\documentclass[sigconf, authorversion,nonacm]{acmart}
\AtBeginDocument{%
  \providecommand\BibTeX{{%
    \normalfont B\kern-0.5em{\scshape i\kern-0.25em b}\kern-0.8em\TeX}}}

\usepackage{bm} 
\usepackage{hyperref}
\usepackage{enumitem}   
\usepackage{mdwlist}    
\usepackage{outlines}   
\usepackage{fancyhdr}
\usepackage{multirow} 
\usepackage{array}
\newcolumntype{C}[1]{>{\centering\arraybackslash}p{#1}}

\usepackage{amssymb}
\usepackage{bbm}
\usepackage{booktabs} 
\usepackage{amsmath,amsfonts}

\usepackage{graphicx}
\usepackage{textcomp}
\usepackage{subcaption}

\theoremstyle{definition}

\usepackage[multiple]{footmisc}

\definecolor{myKeywords}{HTML}{CC3333}  
\definecolor{lightblue}{RGB}{102,153,255}
\definecolor{forestgreen}{RGB}{34,139,34}

\usepackage{makecell}

\definecolor{tabblue}{rgb}{0.12156862745098039, 0.4666666666666667, 0.7058823529411765}      
\definecolor{tabmyorange}{rgb}{1.0, 0.4980392156862745, 0.054901960784313725}                   
\definecolor{tabgreen}{rgb}{0.17254901960784313, 0.6274509803921569, 0.17254901960784313} 
\definecolor{tabpurple}{rgb}{0.5803921568627451, 0.403921568627451, 0.7411764705882353}         
\definecolor{tabred}{rgb}{0.8392156862745098, 0.15294117647058825, 0.1568627450980392}    
\definecolor{tabbrown}{rgb}{0.5490196078431373, 0.33725490196078434, 0.29411764705882354}   

\usepackage[ruled,vlined]{algorithm2e} 
\usepackage{tikz}

\usepackage{stmaryrd}
\usepackage{listings}

\begin{document}


\title{AutoIndexer: A Reinforcement Learning-Enhanced Index Advisor Towards Scaling Workloads}

\author{Taiyi Wang}
\affiliation{%
  \institution{University of Cambridge}
  \city{Cambridge}
  \country{United Kingdom}
}
\email{Taiyi.Wang@cl.cam.ac.uk}

\author{Eiko Yoneki}
\affiliation{%
  \institution{University of Cambridge}
  \city{Cambridge}
  \country{United Kingdom}
}
\email{eiko.yoneki@cl.cam.ac.uk}
\authornote{Corresponding author.}

\begin{abstract}

Efficiently selecting indexes is fundamental to database performance optimization, particularly for systems handling large-scale analytical workloads. While deep reinforcement learning (DRL) has shown promise in automating index selection through its ability to learn from experience, few works address how these RL-based index advisors can adapt to scaling workloads due to exponentially growing action spaces and heavy trial-and-errors. To address these challenges, we introduce \emph{AutoIndexer}, a framework that combines workload compression, query optimization, and specialized RL models to scale index selection effectively. By operating on compressed workloads, \emph{AutoIndexer} substantially lowers search complexity without sacrificing much index quality. Extensive evaluations show that it reduces end-to-end query execution time by up to 95\% versus non-indexed baselines. On average, it outperforms state-of-the-art RL-based index advisors by approximately 20\% in workload cost savings while cutting tuning time by over 50\%. These results affirm \emph{AutoIndexer}’s practicality for large, diverse workloads.

\end{abstract}

\maketitle


\section{Introduction}






Modern database systems heavily rely on index advisors to optimize query performance, yet these tools face mounting challenges as workload complexity grows. Index selection—a critical component of physical database design—aims to identify an optimal set of indexes that minimize query execution costs while respecting system constraints~\cite{schlosser2019efficient,kossmann2022swirl}. This is particularly crucial for SQL databases handling OLAP workloads, where complex analytical queries often involve large-scale data processing and multiple table joins~\cite{siddiqui2024ml}. In cloud-based business applications, we increasingly encounter \textbf{scaling workloads}—scenarios where thousands of queries must be processed within a single tuning window or request. Effectively handling these large-scale workloads is crucial, as it directly impacts system performance, resource utilization, and operational costs in cloud environments~\cite{brucato2024wred,siddiqui2022isum}.

\begin{figure}[ht]
    \centering
    \includegraphics[width=\linewidth]{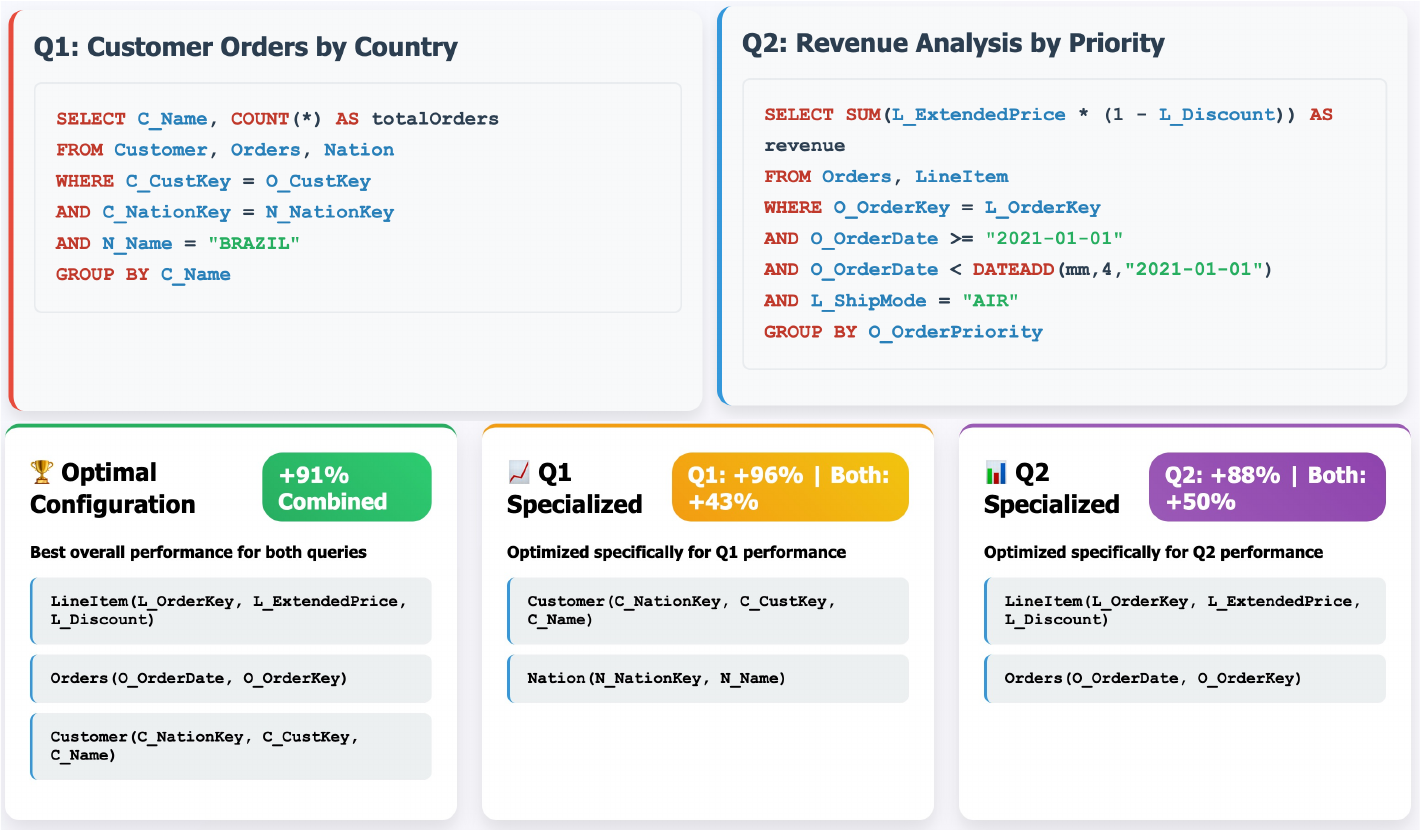}
    \caption{Showcase of the critical role of effective index advisors in joint index selection under representative TPC-H workload.}
    \label{fig:example}
    \vspace{-5pt}
\end{figure}

\begin{figure*}[ht]
\centering
\includegraphics[width=\linewidth]{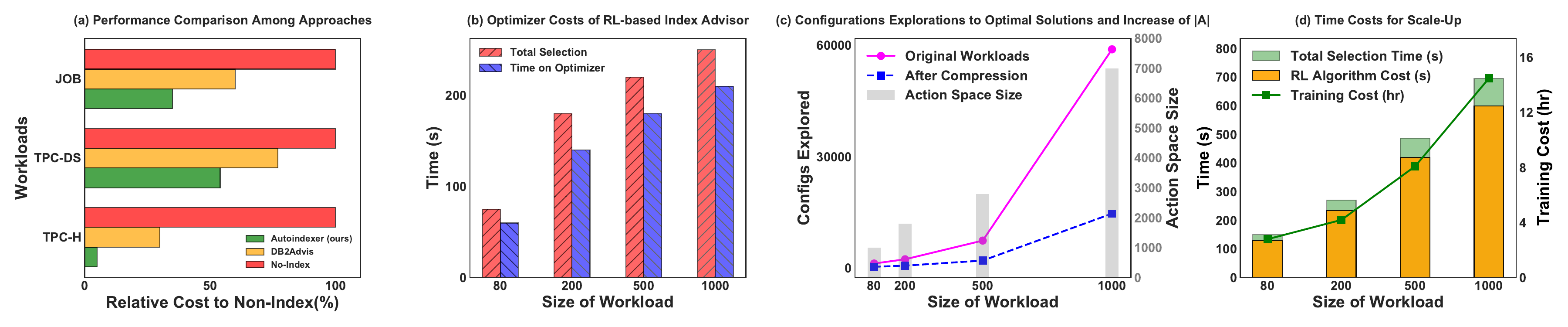}
    \vspace{-15pt}
\caption{Motivation and challenges in scaling-aware index selection (Workload size denotes the number of queries in a single tuning request):
(a) Workload runtime improvements achieved by \emph{AutoIndexer} and DB2Advis~\cite{valentin2000db2} compared to no-index baseline; 
(b) Growing optimizer call (requests of cost-model estimations) overhead in the \emph{SOTA} RL-based approach as TPC-DS workload size increases, demonstrating escalating trial-and-error costs; 
(c) Growth in configuration space exploration required for optimal solutions by the \emph{SOTA} RL-based approach; 
(d) Time cost breakdown of the \emph{SOTA} RL-based approach on JOB benchmark showing rising selection and training overhead (in hours) for larger workloads, storage budget=5GB.}
\label{Perf}
\vspace{-5pt}
\end{figure*}



To illustrate these challenges, consider the TPC-H example in Figure~\ref{fig:example}, where tuning for both queries $Q_1$ and $Q_2$ yields a 91\% improvement, yet isolating either query cuts the gain to 50 \% ($Q_2$ only) or 43 \% ($Q_1$ only).  The gap stems from \emph{shared} multi-attribute indexes: seemingly independent queries compete for the same physical design, and the number of such index-column combinations grows in a combinatorial way, rendering grid or random search ineffective within realistic time budgets. The observations below further motivate our work and highlight current gaps in handling such interactions:

\paragraph{O1: Index Advisor is Critical.}
Figure~\ref{Perf}(a) demonstrates substantial performance gains over no-index baselines across multiple PostgreSQL benchmarks, including heuristic-based \emph{DB2Admin}~\cite{valentin2000db2} and our own RL advisor. This underscores how intelligent index selection—particularly for multi-attribute indexes on OLAP queries~\cite{lan2020index,zhou2024breaking}—can significantly reduce runtimes amid combinatorial complexity.
\paragraph{O2: RL-enhanced Index Advisor Shows Promise.} 

Recent works~\cite{kossmann2022swirl,sadri2020drlindex,wang2024ia2,lan2020index} have reached consensus that RL-based approaches offer compelling advantages over traditional heuristic methods by recasting the combinatorial index selection problem as a sequential decision-making task($\S$~\ref{sec:bg}). Unlike conventional additive or reductive strategies that become computationally prohibitive with large workloads, RL methods can deploy learned policies for rapid index selection decisions. Once trained, given normal-size workloads which contains tens to hundreds of queries, these models can make near-optimal choices without expensive \emph{"what-if"} calls during runtime, leading to both faster model deployment and better solutions. In contrast, traditional index advisors~\cite{chaudhuri1997efficient,schlosser2019efficient,valentin2000db2} require minutes to hours searching within large configuration spaces (but just save seconds-minutes as outcome). As shown in Figure~\ref{Perf}(a), we compare our RL-based method against DB2Advis~\cite{valentin2000db2} on established workloads (TPC-H, TPC-DS, and JOB) in our preliminary experiments. DB2Advis was selected as our baseline because prior works~\cite{kossmann2022swirl,lan2020index} identified it as one of the most advantageous and representative traditional methods. Our results demonstrate that the RL-based approach achieves superior solution quality in terms of workload execution time reduction. Although comprehensive empirical validation across all prior RL-based approaches~\cite{kossmann2022swirl,wang2024ia2,lan2020index,sadri2020drlindex} and traditional methods~\cite{valentin2000db2,chaudhuri1997efficient,schlosser2019efficient,idreos2009self} remains an open challenge, these initial studies demonstrate the promising potential of RL-based index selection in both efficiency and solution quality.

\paragraph{O3: Scaling Workloads Expose RL Bottlenecks.}
As workloads enlarge (with more queries per tuning window), RL advisers encounter two fundamental limits (see baseline \emph{SWIRL} in Figures~\ref{Perf}(b)–(d) and \cite{kossmann2022swirl}).  \textit{(i) “What-if’’ cost inflation}: RL training involves extensive trial-and-error to estimate potential index benefits, and the resulting optimizer calls can dominate overall tuning time (Figure~\ref{Perf}(b)). As queries and columns increase, each additional RL step demands more cost estimations, inflating total overhead. \textit{(ii) Disaster of Exploded Action Space (DEAS)}: the action set—a candidate index for every key combination—grows combinatorially ($\binom{5}{1} + \binom{5}{2} + \binom{5}{3} = 25$ for five attributes; real schemas can easily reach above \(10^{4}\)) \cite{lan2020index,wang2024ia2}.  Figure~\ref{Perf}(c) shows how 3-column keys on TPC-DS blow up configuration count; Figure~\ref{Perf}(d) illustrates training/selection times stretching into hours.  These effects complicate the RL agent’s exploration–exploitation trade-off, driving up both computation time and the risk of failing to converge on high-quality indexes. Despite these challenges, very few works have explored RL-specific optimizations for large-scale index advisors until now.

\begin{figure}[htbp]
\centering
    \includegraphics[width=1.0\linewidth]{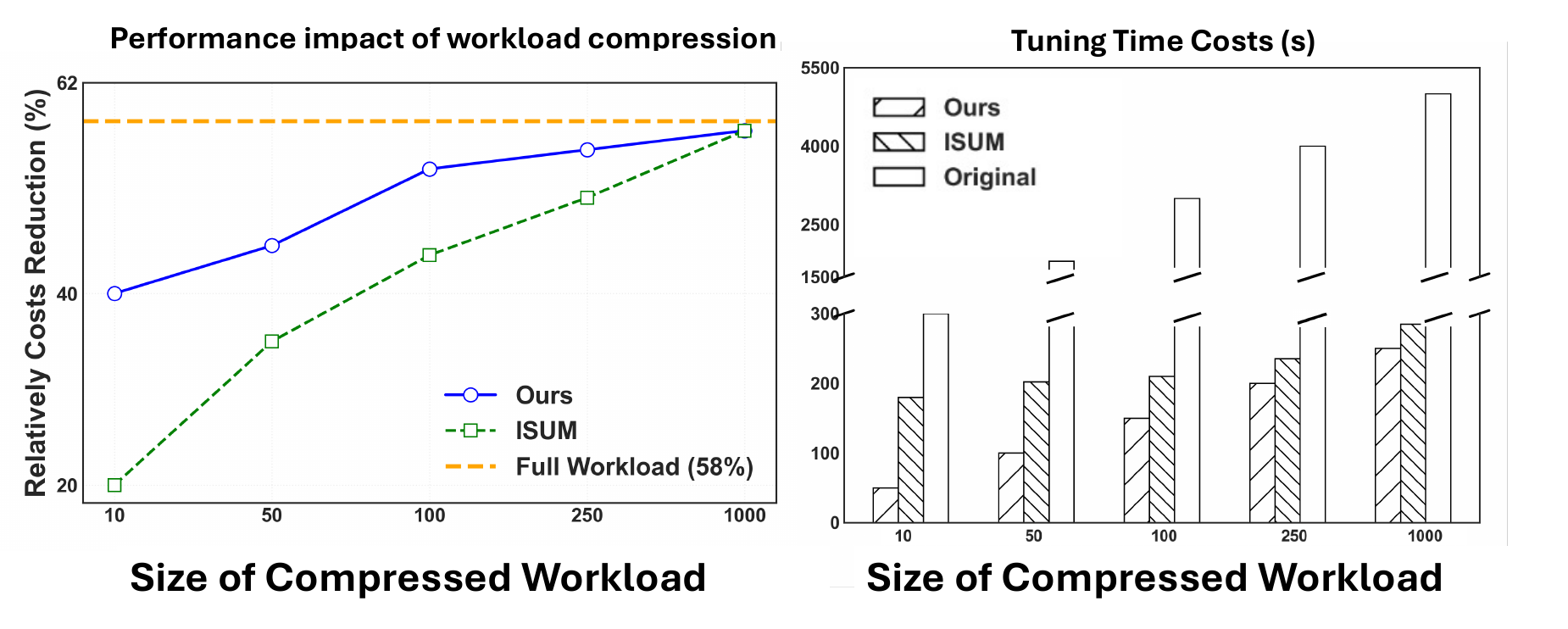}
\caption{Performance comparisons showing the impact of workload compression on index selection for TPC-DS, where \textsc{Isum} compression is integrated with our RL-based approach.}
\label{fig:compress}
\end{figure}

\paragraph{O4: Workload Compression Facilitates Scalable Indexing.}
To address the escalating cost of iterative optimizer calls and unbounded index candidates, prior studies have advocated \emph{workload compression}~\cite{siddiqui2022isum,brucato2024wred}, which merges or prunes queries while preserving near-optimal index quality. By trimming thousands of queries to only a few hundred, compression reduces both the number of columns considered and the volume of cost estimation calls (\emph{"what-if"} calls) that RL must execute. As demonstrated in Figure~\ref{Perf}(c), shrinking the workload constrains the RL agent's necessary explorations, which lowers the agent's trial-and-error overhead. Our experiments show that this approach effectively balances efficiency and quality across multiple dimensions. Figure~\ref{fig:compress}(a) demonstrates that index selection on compressed workloads achieves comparable cost reductions to tuning on original workloads, with the 58\% representing runtime reduction obtained directly from indexing full workloads. Notably, given equivalent compression sizes, our approach surpasses \textsc{Isum}~\cite{siddiqui2022isum} by effectively integrating compression with RL-based action-space optimization, demonstrating superior performance in maintaining solution quality while reducing computational overhead. Meanwhile, Figure~\ref{fig:compress}(b) reveals substantial savings in index selection time, further validating the efficiency gains of our compressed workload approach.


\paragraph{O5: Existing Compression Methods Fall Short for RL} 
Despite these benefits, the performance gap between existing query-centric compressors and our approach in Figure~\ref{fig:compress}(a) and Figure~\ref{fig:compress}(b) highlights how query-level compression techniques, e.g., \textsc{Isum}~\cite{siddiqui2022isum}, fail to address the \emph{DEAS} faced by RL-based advisors. \textsc{Isum}~\cite{siddiqui2022isum} removes entire queries by estimating each query's benefit in lowering optimizer-call overhead, rather than evaluating column-specific impacts on indexing. This approach risks discarding columns essential for RL's action-space pruning and introduces overhead through repeated pairwise query comparisons. Similarly, Another recent compressor \textsc{WRED}~\cite{brucato2024wred} employs statistic-based modeling with AST transformations but operates through query-level heuristics that complicate integration with RL-based advisors, which naturally function at column granularity. Meanwhile, \textsc{WRED}'s complex AST traversals are difficult to reproduce and often rely on strong expert experiences\footnote{Note that Figure~\ref{fig:compress} focuses on comparing with \textsc{Isum} as it represents the current state-of-the-art in workload compression with available implementations, while \textsc{WRED}'s closed-source nature and reproduction challenges limit direct experimental comparison.}. These existing compression methods rely on numerous heuristics for table/subquery connections, making them less robust to schema changes and poorly suited for RL-based advisors that operate at column granularity. The solution requires balancing accuracy and efficiency while directly mapping to the RL action space to enhance indexing performance.


Based on these insights, we introduce \emph{AutoIndexer}, a novel framework that systematically addresses the challenges of large-scale index selection through several key contributions:

(i) A novel workload compression technique using weighted-bipartite graphs and heuristic rules to efficiently merge redundant queries while preserving performance characteristics.

(ii) Enhanced RL models with semantic embeddings and adaptive action masking that capture workload-index relationships to enable efficient exploration of the index selection space.

(iii) Comprehensive experimental evaluation across standard benchmarks demonstrating significant improvements in both efficiency and quality compared to existing approaches.


Overall, to the best of our knowledge, \textbf{\emph{AutoIndexer} is the first index adviser that \emph{co-designs} scaling-aware workload compression with a representation-driven, action-optimized RL engine, tightly coupling system-level mechanisms with algorithmic innovation to tame the combinatorial explosion of index choices towards OLAP workloads.}



The rest of this paper is organized as follows:
Section~\ref{sec:bg} introduces the problem formulation and key RL concepts;
Section~\ref{sec:autoindexer} describes the \emph{AutoIndexer} system workflow;
Section~\ref{sec:experiments} discusses our experimental designs and findings; Section~\ref{sec:related} overviews related work; and finally, Section~\ref{sec:conclusion} concludes the paper, and acknowledgements and artifact announcement follow.

\section{Background}
\label{sec:bg}

\subsection{Index Selection Problem (ISP)}
\label{sec:ISP}

The \emph{Index Selection Problem (ISP)} entails choosing an optimal subset of indexes, \(I^* \subseteq I\), to minimize the execution cost of a workload \(W\) under a storage budget \(C_{\max}\) and a maximum index width \(W_{\max}\):
\begin{equation*}
I^* \;=\; \arg \min_{I' \subseteq I} \text{Cost}(W, I')
\quad 
\text{s.t.}
\quad 
C(I') \;\le\; C_{\max}
\quad 
\end{equation*}
where each chosen index \(i \in I'\) must have \(\omega(i) \le W_{\max}\). Although query optimizers can estimate \(\text{Cost}(W,I')\) via hypothetical indexes~\cite{HypoPG}, the ISP remains challenging due to its massive configuration space (especially for multi-attribute indexes), strong interdependencies among indexes, and the overhead of evaluating numerous \emph{"what-if"} scenarios~\cite{valentin2000db2,brucato2024wred}. 

\subsection{Reinforcement Learning for ISP}
\label{sec:rl}

Reinforcement Learning (RL) approaches the Index Selection Problem (ISP) by learning a mapping between \emph{states} \(s_t\) (representing current index configuration and workload context) and \emph{actions} \(a_t\) (such as adding or removing indexes) to minimize execution cost~\cite{sutton2018reinforcement}, where t denotes the current timestep in the sequential decision process. After each action, the environment transitions to a new state \(s_{t+1}\) with a \emph{reward} \(r_t\) reflecting workload performance improvement. The objective is to learn a policy \(\pi_\phi\) that maximizes the discounted return \(\sum_{\tau=t}^T \gamma^{(\tau - t)} \, r_\tau\) while respecting budget constraints \(C(I_t) \leq C_{\max}\), where T represents the final timestep of the episode. For each episode, the index selection process terminates when the specified memory budget is exhausted and then a new exploration episode will begin. Modern policy gradient methods~\cite{schulman2017proximal} enable direct learning of policy parameters \(\phi\), transforming states into action distributions (\(a_t \sim \pi_\phi(s_t)\)) without exhaustively exploring all configurations. As a result, a trained RL agent can make sequential decisions to optimize index configurations until reaching specified budget constraints (e.g., storage limits for indexes in our context).

This framing renders index selection as a specialized \emph{combinatorial optimization problem}, where the agent incrementally constructs near-optimal index sets through strategic exploration and exploitation. However, the exponential complexity of large-scale workloads presents significant challenges, particularly in handling the \emph{DEAS}, i.e., combinatorial explosion of possible configurations and complex constraints that invalidate many potential actions. Detailed formulations are presented in $\S$\ref{sec:RL_selection}.

\vspace{-5pt}
\section{AutoIndexer: Scaling Workload-Aware Index Advisor Using RL}
\label{sec:autoindexer}
\emph{AutoIndexer} orchestrates three stages to efficiently tune large workloads, as depicted in Figure~\ref{fig:system}:

\begin{figure*}[ht]
\centering
    \includegraphics[width=\linewidth]{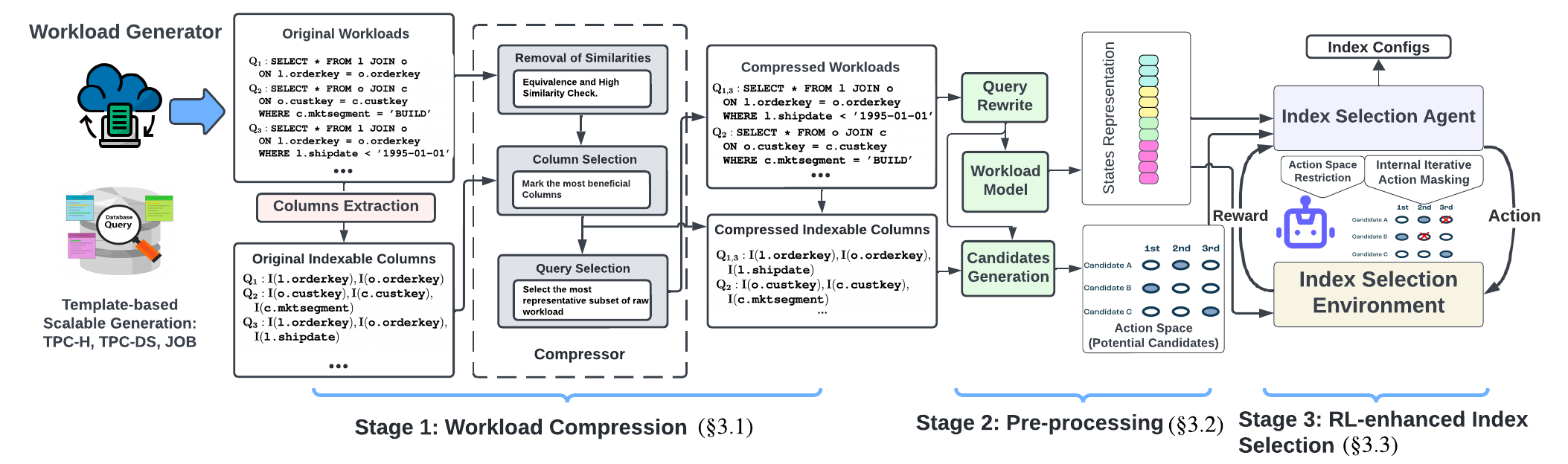}
\caption{System pipeline of \emph{AutoIndexer}}
\label{fig:system}
\end{figure*}

\textbf{Compression (\S\ref{sec:comp}).} Our first stage tackles large workloads (often thousands of queries) via a \emph{column-centric} compression strategy. Unlike query-centric methods~\cite{brucato2024wred,siddiqui2022isum}, we focus on the \emph{value} of remaining attributes, discarding both columns and entire queries that lack meaningful indexing opportunities. By combining column-centric awareness with query pruning, we effectively shrink the candidate index space while preserving the most impactful structures for downstream selection.

\textbf{Preprocessing (\S\ref{sec:pre}).} After compression, we first apply query rewrites on the compressed workload to boost performance. We then systematically generate potential indexes from the pruned query set and capture essential query features through learned workload representations as our input to the downstream RL model. By operating on the compressed workload only, we eliminate unnecessary computation for discarded columns.

\textbf{Index Selection with RL (\S\ref{sec:RL_selection}).}
Finally, we frame index selection as a Markov Decision Process. During the \emph{training phase}, our State-wise RL agent repeatedly explores index configurations-encoded as states reflecting the current indexes, resource usage, and workload characteristics—and updates its policy based on feedback from the environment. Once trained, the resulting policy is \emph{deployed} to rapidly pick or discard index candidates for new workloads under similar constraints. Crucially, operating on the compressed workload and preprocessed features reduces the action space, enabling both faster training convergence and stable online performance without compromising recommendation quality.

In the following sections, we detail the technical components and innovations of each stage.

\subsection{Compression Stage}
\label{sec:comp}

Modern database systems employ index advisors to recommend optimal index configurations. Given the resource-intensive nature of index selection for large workloads, workload compression techniques are crucial for scalable index tuning.


\subsubsection{Targets and Motivation}
Workload compression addresses the scalability challenges in index tuning by reducing large query workloads to their essential components while preserving key characteristics for downstream optimization. While existing index advisors~\cite{chaudhuri1997efficient, valentin2000db2, schlosser2019efficient} can identify optimal configurations, they rely heavily on manual tuning of compression parameters, including similarity thresholds and target workload sizes. This dependence on expert-defined heuristics limits their adaptability and practicality in dynamic database environments.

The core challenge lies in automating this compression process: how to efficiently remove redundant queries and columns while minimizing the impact on final index quality. To achieve this goal, we develop a novel framework that combines targeted heuristic rules with an integer linear programming (ILP) formulation. This hybrid approach automatically identifies a minimal yet representative set of queries and columns that can effectively guide the index advisor toward near-optimal solutions, eliminating the need for manual parameter tuning while preserving essential workload characteristics.


\subsubsection{Compression Steps}
Our workload compression proceeds in three phases: 
\emph{(i)~similarity and equivalence removal}, 
\emph{(ii)~column selection}, and
\emph{(iii)~query selection}.
Each phase incrementally filters or transforms the workload to remove redundancies and emphasize index-relevant structures.

\paragraph{Phase 1: Similarity and Equivalence Removal.}
To effectively reduce the action space for downstream RL, we first identify and consolidate similar and equivalent queries\footnote{Equivalence is verified by the rewrite rules and validation methods in Section~\ref{sec:pre}.}. When duplicates or near-duplicates are found, we preserve the query whose set of columns appears more frequently across the workload, based on aggregated usage statistics. We also optimize the workload by eliminating queries that can be fully rewritten into an existing one using abstract syntax tree transformations and equivalence rules. To measure query similarity, we use a Jaccard coefficient~\cite{niwattanakul2013using} over sets of indexable columns:
\[
Sim(q_i, q_j) \;=\; \frac{C_{q_i}\,\cap\,C_{q_j}}{C_{q_i}\,\cup\,C_{q_j}},
\]
where \(C_{q_i}\) denotes the set of indexable columns in query \(q_i\). Queries surpassing a certain similarity threshold are merged into a single representative, thereby minimizing duplicate effort in the subsequent index tuning phase.

\paragraph{Phase 2: Column Ranking and Selection.}

To enhance column-wise awareness, we identify columns most likely to improve performance by computing a benefit score for each column. Our approach first evaluates potential performance gains using \emph{"what-if"} cost estimations on the \emph{compressed} workload. Although these estimator calls still involve optimizer queries, they are far more cost-effective than allowing a full-scale RL approach to repeatedly evaluate every query–column pairing across numerous training episodes in the original, uncompressed workload. The \emph{ColBenefit} for a column c is computed as the average cost reduction ratio across all queries q that reference c:
\[
\text{ColBenefit}(c) = \frac{1}{|Q_c|}\sum_{q \in Q_c} \frac{Cost(q) - Cost(q|c)}{Cost(q)},
\]
where $Q_c$ is the set of queries referencing column c, Cost(q) is the original query cost, and Cost(q|c) is the estimated cost with an index on c. We then compute the final score as:
\[
\text{Score}(c) \;=\; D(t)\,\times\,\text{\emph{ColBenefit}}(c),
\]
where \(D(t)\) represents the importance of table \(t\) based on its size and query reference frequency:
\(
D(t) \;=\; \lvert t \rvert \;\times\; N_{ref}(t,W).
\)
Here, \(\lvert t\rvert\) denotes the table's row count and \(N_{ref}(t,W)\) counts queries referencing table \(t\). This scoring reflects two key observations:

\noindent\textbf{Observation~1:} \emph{"Tables with more rows gain greater benefit from indexing."}

\noindent\textbf{Observation~2:} \emph{"Indexes are more beneficial on tables that are accessed more frequently."}

By multiplying row count (\(\lvert t\rvert\)) and reference frequency \(N_{ref}(t,W)\) into \(D(t)\), we prioritize columns on large tables or heavily used tables—precisely where an index is most beneficial. We retain the top 85\% of columns based on their scores, effectively reducing the search space while preserving the most impactful candidates for indexing.

\paragraph{Phase 3: Selecting Representative Queries via a Weighted Bipartite Graph}
\label{sec:phase3}

Phase~3 distills the (still-redundant) workload emerging from the previous phases into a compact, high-value subset of queries.  
The central idea is to cast the interaction between queries and indexable columns as a \emph{weighted bipartite graph} and then
identify—through a small integer program—the set of queries that maximises overall benefit while still covering the columns that matter.

\textbf{Notation.}
Let  
\(
\mathcal{Q}=\{q_1,\dots,q_m\}
\) denote the remaining workload after Phase~2 and  
\(
\mathcal{C}=\{c_1,\dots,c_n\}
\) the universe of columns that can be indexed.
For brevity we write \(E\subseteq \mathcal{Q}\times\mathcal{C}\) for the set of query–column references
and use binary variables \(x_i\) and \(y_j\) to indicate whether query \(q_i\) or column \(c_j\) is selected, respectively.

\textbf{Step~1: Graph construction.}
We create a bipartite graph \(G=(\mathcal{Q}\cup\mathcal{C},E)\) with  
an edge \((q_i,c_j)\in E\) whenever query \(q_i\) mentions column \(c_j\).
This graph compactly records the coverage dependencies between candidate queries and physical design choices (indexes).

\textbf{Step~2: Edge weighting.}
Each edge is annotated with a weight that balances \emph{individual benefit} and \emph{synergy} with similar queries:
\[
w_{i,j}= \underbrace{\Delta\!\operatorname{Cost}(q_i,c_j)}_{\text{direct estimated speed-up}}
          +\;
          \lambda\,f_i
          \Bigl(
            \sum_{k\neq i}
              \operatorname{Sim}(q_i,q_k)\,
              \Delta\!\operatorname{Cost}(q_k,c_j)
          \Bigr).
\]
\begin{itemize}[leftmargin=*]
  \item \(\Delta\!\operatorname{Cost}(q_i,c_j)\) estimates
        the runtime cost improvement of \(q_i\) if \(c_j\) were indexed.  
  \item \(f_i\) is the observed frequency of \(q_i\).  
  \item \(\operatorname{Sim}(q_i,q_k)\) is the overlap score already computed in Phase~1.  
  \item \(\lambda\) tunes the relative importance of synergistic gains.
\end{itemize}

\textbf{Step~3: ILP-based subset selection.}
We then solve a small Integer Linear Program whose objective is to maximise the total collected edge weight while guaranteeing that every \emph{critical} column is still covered:
\[
\begin{aligned}
\text{Maximise}\quad & \sum_{(i,j)\in E} w_{i,j}\,x_i\,y_j \\[2pt]
\text{subject to}\quad &
   \sum_{(i,j)\in E} x_i\,y_j \;\ge\; \textsc{Cover}(E), \\[2pt]
& x_i\in\{0,1\},\; y_j\in\{0,1\}\quad \forall i,j.
\end{aligned}
\]
\(\textsc{Cover}(E)\) is a tunable lower bound on the number of distinct query–column pairs that must remain.  
The decision variables \(y_j\) merely ensure coverage; our final output is the set of queries with \(x_i=1\). We denote the compressed workload as \(\widehat{\mathcal{Q}}\) for the subsequent use.

\begin{figure}
    \centering
    \includegraphics[width=\linewidth]{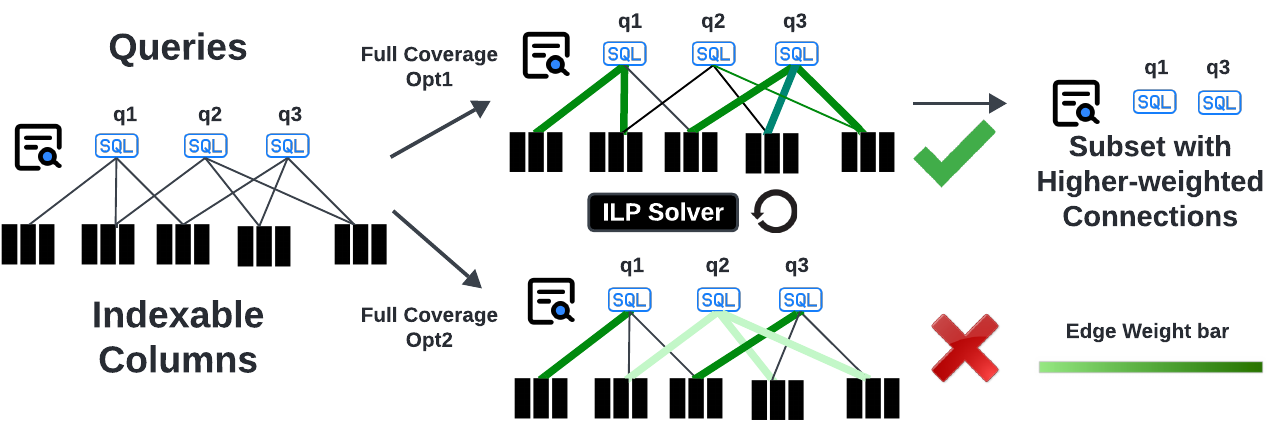}
    \caption{Query selection via weighted bipartite graph}
    \label{fig:bp_comp}
\end{figure}

Most compression techniques fix a target ratio or heavily depend on template hyper-parameters~\cite{siddiqui2022isum,brucato2024wred}.  
By contrast, our graph view lets the ILP decide \emph{how many} queries to retain purely on the basis of expected benefit, thereby adapting automatically to workload skews and inter-query similarities. Figure~\ref{fig:bp_comp} visualizes the bipartite instance for a toy workload. Edges with high weights (thick lines) steer the ILP towards a solution that keeps only the queries whose collective presence maximises performance uplift while discarding low-impact or redundant ones.

\paragraph{Hyperparameter Configuration and System Optimization.}
Our compression module selects its few critical hyperparameters automatically: an empirically derived 85\% column-ranking threshold (via Pareto analysis) captures almost all performance gains without incurring extra overhead, while delta-cost values and balance coefficients are seeded from optimizer cost estimates and refined with a shallow grid search. 
This graph-based pruning eliminates the extensive traversal, structure, and rewriting knobs of \textsc{WRED}~\cite{brucato2024wred} and the fixed heuristics of \textsc{Isum}~\cite{siddiqui2022isum}, cutting search complexity with negligible loss of index quality.

Because jointly tuning compression and RL components creates an intractably large hyperparameter space, true end-to-end optimization is impractical; instead, we combine ablation studies with component-level grid search, achieving near-optimal joint performance at a small fraction of the computational cost.

\subsection{Pre-processing Stage}
\label{sec:pre}

After obtaining the effectively compressed workloads, we establish a robust foundation for downstream index selection through three key pre-processing components: query rewrite, candidate enumeration, and workload modeling.

\subsubsection{SMT-Guided Query Rewrite}
\label{sec:phase4}

Once we have compressed the original workload into the high-value subset
\(\widehat{\mathcal{Q}}\), we apply a \emph{provably sound} rewrite pass that prepares these queries for the downstream RL-based index-selection and candidate-generation stages.  The pass ports \textsc{WeTune}’s SMT (Satisfiability Modulo Theories) methodology~\cite{wang2022wetune,zhou2021learned} into our pipeline. The goal is to discover a series of efficient and equivalent rewrite rules.


\paragraph{Step 1 — Template enumeration and U-expression encoding.}
For each query in \(\widehat{\mathcal{Q}}\) we enumerate "interesting" patterns (nested subqueries, redundant \texttt{GROUP BY}s, non-canonical predicates) and
translate every candidate fragment into a \emph{U-expression}~\cite{chu2018axiomatic}:
\[
\llbracket\mathcal{Q}\rrbracket : \mathit{Tuple}\;\longrightarrow\;\mathbb{N},
\qquad
t \;\mapsto\; \text{multiplicity of } t .
\]
This algebraic, bag-semantics representation makes equivalence proofs tractable. Table~\ref{tab:u-ops_0} summarizes the SQL–to–U-expression mapping.

\begin{table}[t]
\captionsetup{font=small}
\centering
\footnotesize
\caption{Representative U-expression Operators Corresponding to Common SQL Elements}
\label{tab:u-ops_0}
\begin{tabular}{p{2.5cm}p{5cm}}
\toprule
\textbf{SQL Operator} & \textbf{U-expression Form} \\
\midrule
Projection (\texttt{SELECT}) 
 & $\text{Proj}(f)(t) = f(t)$ where $f$ is the projected subset \\
Join (\texttt{JOIN}) 
 & $\text{Join}(f_1, f_2)(t) = \sum_{t_1 \bowtie t_2 = t} f_1(t_1) \cdot f_2(t_2)$ \\
Selection (\texttt{WHERE}) 
 & $\text{Sel}_{\theta}(f)(t) = f(t) \cdot \delta(\theta(t))$ \\
Union / Set Op 
 & $\text{Union}(f_1,f_2)(t) = f_1(t) + f_2(t)$ \\
Grouping / Aggregation 
 & $\text{Group}(f)(g) = \cdots$ (maps tuples to group-keys, etc.) \\
\bottomrule
\end{tabular}
\end{table}

\paragraph{Step 2 — Constraint synthesis.}
Given a source–destination pair
\((\mathcal{Q}_{\text{src}},\mathcal{Q}_{\text{dest}})\),
we generate the \emph{minimal} logical constraints needed to prove equivalence. These constraints capture relationships between relations, attributes, and predicates:
\begin{gather}
   \text{RelEq}(rel_1, rel_2), \notag \\
   \text{AttrsEq}(attrs_1, attrs_2), \notag \\
   \text{PredEq}(pred_1, pred_2), \notag \\
   \text{NotNull}(rel, attr), \ldots \notag
\end{gather}

where each ensures alignment of logical components (e.g., enforcing that two attributes represent the same value domain). By iterating over different constraint subsets, we generate candidate rule sets, discarding those that are contradictory or fail to improve performance. Constraint sets that are contradictory or do not promise net benefit are pruned immediately.

\paragraph{Step 3 — Equivalence proof via SMT.}
The constraint set \(C\) and the formula
\(
\forall t\,
\bigl(\llbracket\mathcal{Q}_{\text{src}}\rrbracket(t)=
\llbracket\mathcal{Q}_{\text{dest}}\rrbracket(t)\bigr)
\)
are submitted to Z3~\cite{de2008z3}.  An \textsc{unsat} result certifies semantic equivalence; otherwise the candidate rule is rejected.

\paragraph{Step 4 — Ranking and retention.}
Verified rules are ranked by net runtime saving versus rewrite overhead. Transformations that flatten subqueries, prune redundant aggregations, or normalize predicates score highest, whereas low-impact or plan-inflating rules are discarded. Then we will keep these rewrite rules and apply them to optimize queries.

\paragraph{Impact on downstream phases.}
SMT-guided rewrites cannot alter the set of columns that require indexing.  They \emph{can}, however,
simplify each AST and reduce alternative plan space, thereby:
(i) speeding up cost estimation for the RL agent and
(ii) focusing index-candidate generation on genuinely beneficial structures instead
of syntactic clutter.

In summary, this post-compression, SMT-guided rewrite pass delivers a workload that is both semantically faithful and structurally streamlined—ideal input for the RL-based index-selection engine that follows.


\subsubsection{Index Candidates Enumerator} 
Given the \emph{rewritten query set} produced by the SMT-rule–based rewriter, we enumerate potential index candidates in a way that deliberately avoids brute-force explosion.  
First, an \emph{index parser} decomposes each query into predicate, join, and ordering columns and records them in a compressed bitmap matrix.  
A \emph{permutation engine} subsequently re-orders these column subsets while enforcing a user-specified maximum width, \(W_{\max}=5\).

To keep the candidate space tractable, three sequential filters are applied.  
(i) \emph{Relevance scoring} ranks columns by a weighted mixture of predicate selectivity, join frequency, and their appearance in \textsc{group by} or \textsc{order by} clauses.  
(ii) A \emph{validation check} discards permutations that violate functional dependencies, involve unsupported data types, or conflict with system constraints.  
(iii) Finally, a \emph{heuristic pruning} phase—adapted from the insights of Lan\,\textit{et al.}~\cite{lan2020index}—eliminates any permutation that is a strict superset of a validated candidate whose marginal gain falls below a predefined threshold, or whose estimated maintenance overhead would exceed the storage budget.  
The remaining permutations constitute the \emph{Index Candidate Pool} delivered to the RL agent, ensuring that only high-impact, feasible indexes participate in subsequent decision making.

\subsubsection{Workload Model} 
\begin{figure}
    \centering
    \includegraphics[width=\linewidth, height=0.54\linewidth]{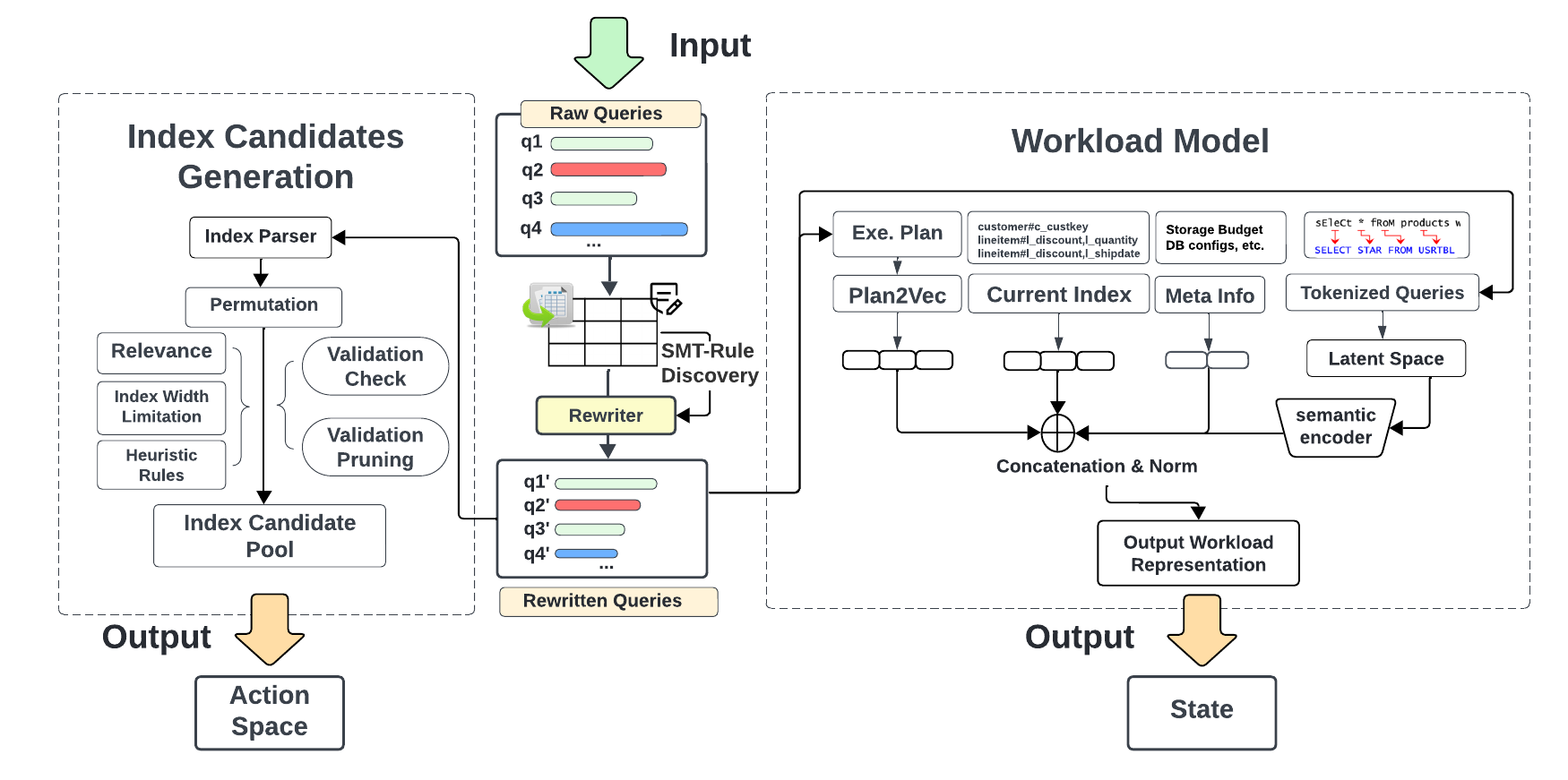}
    \vspace{-20pt}
    \caption{Pre-processing Procedures for State-Wise RL}
    \label{fig:WM}
\end{figure}

To furnish the RL agent with a rich yet manageable state representation, we construct a \emph{comprehensive workload model} (Figure~\ref{fig:WM}) that concatenates four orthogonal information sources, followed by layer normalization.  
Specifically, the model integrates (i) \emph{execution-plan features} produced by \textsc{Plan2Vec}, which encodes operator types, estimated cardinalities, and optimizer costs; (ii) a sparse bit-vector representing the \emph{current index configuration}; (iii) dense \emph{meta-information} capturing the storage budget, DBMS knob settings, and hardware limits; and (iv) \emph{tokenized query embeddings} obtained from a semantic encoder described below.  
The resulting vector forms the environment state \(s_t\) consumed by the policy network.

In terms of semantic Encoder and plan2vec models, we repurpose the final hidden layer of a RoBERTa-style Transformer that was first pre-trained on generic code–text corpora and subsequently \emph{self-supervised finetuned} on 28 M SQL statements collected from public benchmarks and production traces.  

Finetuning employs two objectives: a \emph{masked-token reconstruction} task that randomly masks SQL keywords, table names, or column names and trains the model to recover them, thereby reinforcing syntactic awareness; and a \emph{contrastive plan-pairing} loss that minimises the embedding distance between queries sharing an identical logical plan but differing in literal constants, thus capturing plan-level semantics independent of bound values.  

At inference time, each query \(q_i\) is passed through the encoder, mean-pooled over statement tokens, and projected to a 256-dim latent space. These embeddings faithfully preserve both syntactic structure and cost-critical patterns—such as the presence of nested-loop joins—thereby enabling effective generalization across previously unseen workloads.



Once these preprocessing steps are complete, the workload is suitably structured for RL-based index selection. In particular, the column and query streamlining performed here ensures that the final compressed workload includes only the most relevant attributes while maintaining essential structural information for the RL agent.

\subsection{Index Selection Stage}
\label{sec:RL_selection}

Rather than relying on standard DRL alone, we propose a tailored RL-based methodology that prunes large action spaces via an adaptive masking mechanism. This design accelerates convergence and enhances solution quality by focusing the agent’s exploration on only the most beneficial index candidates.

\subsubsection{RL Formulation}
Building on Section~\ref{sec:rl}, we pose index selection as a sequential decision process in which an RL \emph{agent} interacts with database environments.~\cite{sutton2018reinforcement,lan2020index}. Key elements within the standard Reinforcement Learning framework are formulated as below: 

\noindent\textbf{State ($s_t$).}  
At timestep \(t\) the environment returns the fixed-length
\emph{workload-model} vector shown in figure~\ref{fig:WM}:  
Plan2Vec embeddings of execution plans, tokenized-query embeddings from the semantic encoder, a bitmap of the current index set, relevant system metadata, and the remaining storage budget.  All components are concatenated and $\ell_2$-normalized.

\noindent\textbf{Action ($a_t$).}  
The discrete action space is the \emph{index-candidate pool} (left side of
figure~\ref{fig:WM}).  Each integer
\(a_t\!\in\!\{0,\dots,|\mathcal{P}|-1\}\) selects one candidate index. Although we target OLAP workloads, we still (i) bundle all \texttt{CREATE INDEX} commands into a single transaction—this grabs the catalogue lock only once and amortizes commit overhead—and (ii) enforce a generous \texttt{statement\_timeout}. Long-running analytic queries are rare but not impossible (e.g., if a join is mis-estimated); without a timeout a single straggler would block the entire vector of parallel roll-outs. Both safeguards add negligible overhead yet keep training throughput predictable.

\noindent\textbf{Reward ($r_t$).}  
To balance performance gain against storage cost, we use the
storage-normalized cost reduction
\begin{equation}
\label{eq:reward}
r_t(I_t^\ast)=
\frac{\bigl(C(I_{t-1}^\ast)-C(I_t^\ast)\bigr)/C(\emptyset)}
     {\bigl(M(I_t^\ast)-M(I_{t-1}^\ast)\bigr)/M(I_{t-1}^\ast)},
\end{equation}
where \(I_t^\ast\) is the index set after executing \(a_t\),
\(C(\cdot)\) is the optimiser’s workload cost estimate, and
\(M(\cdot)\) the cumulative index size.  

During \emph{training}, the RL agent deliberately explores different index choices (some of which may be suboptimal) to discover effective strategies. Once the agent \emph{converges}, it switches to a purely exploitative (deterministic) mode during \emph{deployment}—the trained policy is applied directly to recommend indexes for real workloads. In practice, we found that convergence typically requires on the order of thousands of environment steps, but we reduce the runtime by parallelizing across multiple "Index-Gym" environments.


\subsubsection{Index-Gym: A Vectorized RL Environment.}
\label{sec:index-gym}
To cut wall-clock training time we implement \textbf{Index-Gym}, which provides a light-weight, vectorized test-bed that wraps multiple identical DBMS instances behind the \texttt{gymnasium} \texttt{VectorEnv}~\cite{towers2024gymnasium} API, enabling an RL agent to advance dozens of roll-outs with a single call in distributed settings. Each worker operates in its own container while mounting the same read-only \texttt{base.datadir} through copy-on-write snapshots (overlayfs, ZFS), making \texttt{reset()} operations nearly instantaneous even for multi-GB datasets.  Because every worker exposes the same state \(s_t\) and action space \(a_t\) defined above, an agent can advance all roll-outs with one call:

\begin{lstlisting}[language=Python, basicstyle=\small\ttfamily]
envs = gym.vector.AsyncVectorEnv(
            [lambda: IndexGymEnv(seed=i) 
            for i in range(k)])
agent.learn(total_timesteps = k * 5000)
\end{lstlisting}

\noindent On a 8-core workstation a width of \(k{=}4\) lowers training for the 100-GB TPC-DS workload from roughly 6h to under 2h.\footnote{Further implementation details, hyper-parameters, and convergence diagnostics appear in the supplementary materials.}



\subsubsection{State-Wise RL with Automated Action Pruning}
\label{method_instanceaware}
\begin{figure}
    \centering
    \includegraphics[width=\linewidth]{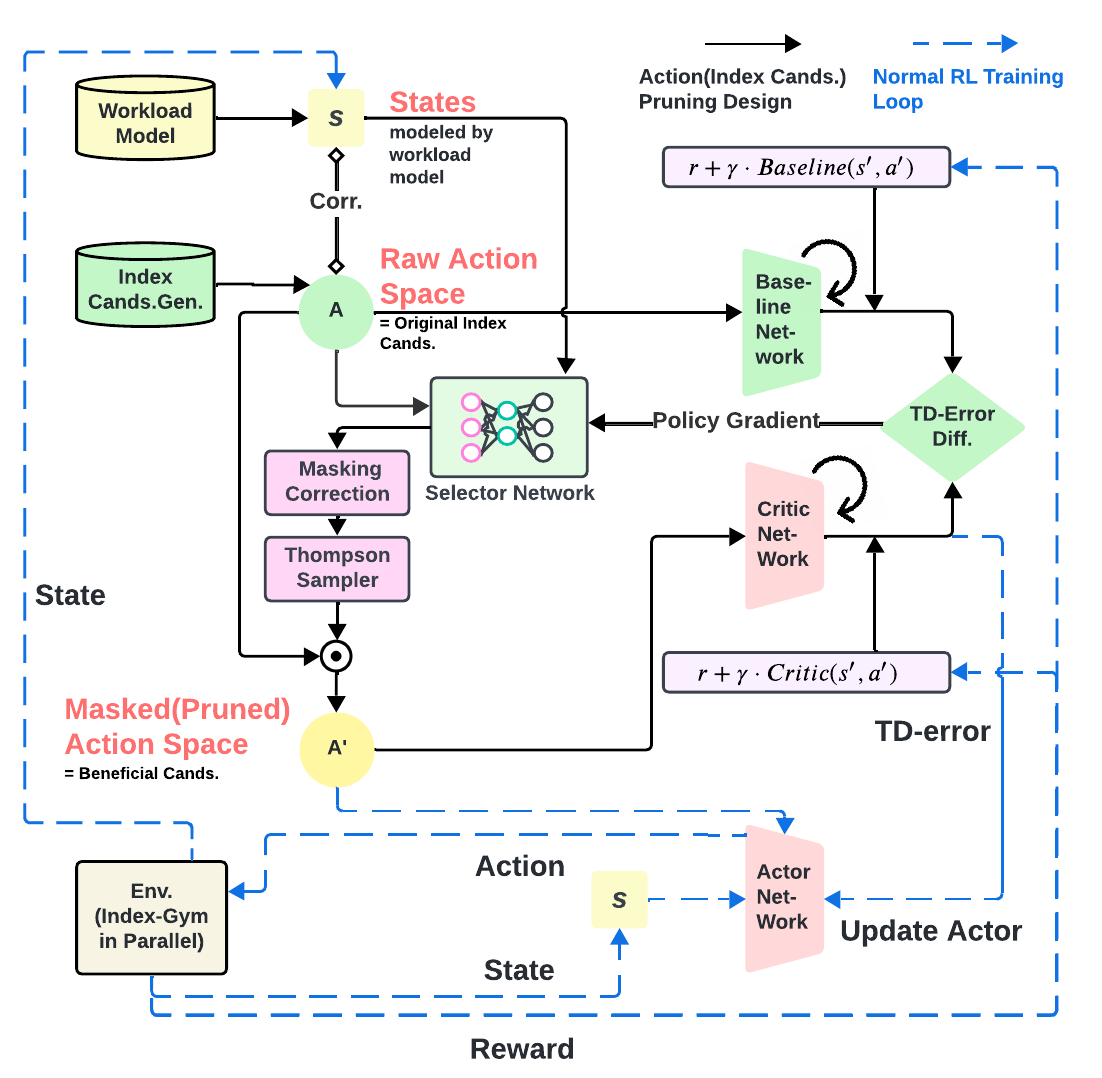}
    \caption{State-wise reinforcement learning framework in AutoIndexer featuring automated action space pruning for index candidates. Areas enclosed by black lines represent the action masking mechanism, while areas bounded by blue dashed lines illustrate the standard reinforcement learning workflow.}
    \label{fig:RL-model}
\end{figure}

Even after query and column pruning, realistic workloads still expose numerous candidate indexes—a severe \emph{Disaster of Exploded Action Space} (DEAS).  Exploring this space exhaustively is infeasible, so we augment a discrete, on-policy actor–critic backbone with a \textbf{state-wise selector network} \(G_{\theta}\) that prunes the action space on the fly (Figure~\ref{fig:RL-model}).

\textbf{Selector-guided masking.}  
For the current state \(s_t\), \(G_{\theta}\) produces a binary mask
\(G_{\theta}(a\!\mid\!s_t)\!\in\!\{0,1\}^{|\mathcal P|}\).  
A Thompson Sampler~\cite{russo2018tutorial} turns these logits into a pruned set
\(\mathcal A'_t\subset\mathcal P\); the actor then selects
\(a_t\!\in\!\mathcal A'_t\).   In parallel, a lightweight \emph{baseline network} estimates the value of the \emph{unmasked} action set, giving a reference against which improvements can be judged.

\textbf{Training dynamics.}  
After executing \(a_t\), the critic computes a TD target \(y'_t\). The selector is updated with a policy-gradient signal derived from the difference between TD errors from masked (fed into the critic network) and unmasked (fed into the baseline network) action space:
\begin{equation}
\label{eq:td-objective}
\min_{\theta,w}\;
\mathbb{E}_{s_i}\!\Bigl[
 \bigl(y'_i -
       Q_{w}\bigl(s_i,
                  a^{(G_{\theta}(a\mid s_i))}\bigr)\bigr)^{2}
 +\lambda\,\lVert G_{\theta}(a\mid s_i)\rVert_{0}
\Bigr],
\end{equation}
where \(\lambda\) enforces sparsity, and $s$ represents the current state including partial indexing decisions and workload statistics. While $a$ enumerates the candidate indexes, $\lambda$ controls the balance between fitting TD errors and maintaining mask sparsity to prune low-value actions. The TD target $y'_i$ accounts for masked or unmasked actions, enabling the selector to learn which actions improve performance.

Intuitively, TD-error improvements earned by the masked policy provide \emph{positive} reinforcement for actions retained by \(G_{\theta}\), while degraded returns penalize them, steering the mask to zero out low-value indexes. The agent operates on a subset of the original workload---pre-selected during column and query pruning---so many low-value actions never appear.  This synergy between compression-based pruning and dynamic masking keeps the environment manageable for large-scale workloads.

\textbf{Distinguishing features.}  Unlike earlier approaches that graft Bernoulli masks onto continuous, off-policy learners such as TD3~\cite{wang2024ia2,sun2022toward}, our method employs a \emph{discrete, on-policy} actor–critic backbone expressly designed for index creation and deletion. The Thompson-guided selector is trained \emph{jointly} with the agent, providing principled uncertainty estimates and accelerating convergence ($\approx$1.7× faster in our experiments).  Because masking follows workload compression, the learner issues far fewer "what-if" cost evaluations while still exploring a curated, high-quality subset of candidate indexes.






\section{Experiments}
\label{sec:experiments}

\subsection{Experimental Setting}
\label{sec:exp_setting}

\paragraph{Implementation and Environment.} 
Our system is implemented in Python using PyTorch, with PostgreSQL 15.6 and HypoPG for index simulation. Experiments run on an 8-core VM with an NVIDIA Quadro RTX8000 GPU. Training is parallelized across four "Index-Gym"-enhanced DBMS environments, each maintaining independent database snapshots for stability.

\paragraph{Benchmark Workloads.}
Table~\ref{tab:workloads} summarizes our evaluation workloads. To ensure fair comparison, we follow established configurations from previous index tuning research~\cite{siddiqui2022isum,brucato2024wred,kossmann2022swirl}. We employ three distinct benchmarks: (1) TPC-H, (2) TPC-DS, and (3) the Join Order Benchmark (JOB) based on IMDB data, which exhibit real-world query patterns and skewed data distributions. For TPC-H and TPC-DS, we use a scale factor of 10. Query instances are generated using the open-source \emph{qgen} utilities for TPC-H and TPC-DS, where we alter column bindings, table references, and parameter values following the official templates~\cite{siddiqui2022isum,stephens2009converting}, then shuffle the resulting queries to simulate diverse, realistic workloads. These workloads collectively provide a comprehensive evaluation across varying query complexities, data distributions, and access patterns.

\begin{table}[htbp]
\captionsetup{font=small}
\centering
\begin{tabular}{@{}lccc@{}} 
\hline
Name & \#Queries & \# Templates & \#Tables \\
\hline
TPC-H ($sf10$) & 2200 & 22 & 8 \\
TPC-DS ($sf10$) & 930 & 93 & 24 \\
JOB & 1120 & 112 & 21 \\
\hline
\end{tabular}
\caption{Summary of evaluation workloads}
\label{tab:workloads}
\vspace{-20pt}
\end{table}

\subsection{Baseline Approaches}
\label{sec:baseline}

We compare \emph{AutoIndexer} against two sets of baselines: \textbf{(i)~Index advisor} and \textbf{(ii)~Workload compressor}.
\paragraph{Index Advisor}
We evaluate three \emph{heuristic} algorithms: 
\emph{AutoAdmin}~\cite{chaudhuri1997efficient}, a classic \emph{cost-driven} additive approach; 
\emph{DB2Advis}~\cite{valentin2000db2}, known as the \emph{fastest} among non-RL solutions\footnote{This is corroborated by evaluations from the open-source  platform proposed in~\cite{kossmann2022swirl}, aligning with the empirical studies in sections 3.1 and 6.1 of that paper}; 
and \emph{Extend}~\cite{schlosser2019efficient}, which is \emph{heuristic-based} and often finds near-optimal solutions.  For \emph{RL-based} baselines, we include \emph{SWIRL}~\cite{kossmann2022swirl}, a \emph{dynamic masking} approach that integrates workload modeling to handle multi-attribute indexes. When sufficiently trained, SWIRL is state-of-the-art for known workloads but can be training-intensive. We also compare against \emph{DRLinda}~\cite{sadri2020drlindex}, which uses a \emph{multi-agent} RL paradigm designed to generalize across unseen query sets with single-attribute index. However, DRLinda lacks native storage-budget constraints; thus, we adapt it by ranking its selected indexes by size and adding them sequentially until the budget is reached, also checking whether smaller remaining indexes can fit. Through these baselines, we capture a diverse set of strategies—spanning conventional additive heuristics to advanced RL formulations—for a comprehensive comparison. 

\paragraph{Workload Compressor}
We likewise compare our proposed compression strategy with \textsc{Isum}~\cite{siddiqui2022isum}, which by default employs rules weighing columns based on query costs, table sizes, and the number of generated index candidates. Although \textsc{WRED}~\cite{brucato2024wred} is another recent technique, it relies on complex query rewrites and transformations that are neither publicly available nor straightforward to reproduce. \textsc{WRED}’s design focuses on merging or restructuring queries, rather than \emph{selecting} a representative subset as in our approach or \textsc{Isum}. Furthermore, \textsc{WRED} applies extensive transformation-based heuristics that may complicate RL-based usage. Consequently, we limit our direct comparison to \textsc{Isum}, which is both open-source and closely aligned with our method’s subset-selection paradigm.


\subsection{Experimental Results}

Our experimental evaluation comprises five key components: (i) end-to-end performance assessment across diverse benchmarks (\S\ref{sec:e2e}), (ii) analysis of training efficiency improvements (\S\ref{sec:training}), (iii) runtime breakdown compared to prior approaches (\S\ref{sec:breakdown}), (iv) systematic evaluation of our different modules through integration with existing methods (\S\ref{sec:RL+Combine}), and (v) ablation studies quantifying each module's contribution (\S\ref{sec:ablation}).

\subsubsection{End-to-End Performance}
\label{sec:e2e}

\begin{figure*}[ht]
\centering
\includegraphics[width=0.95\linewidth]{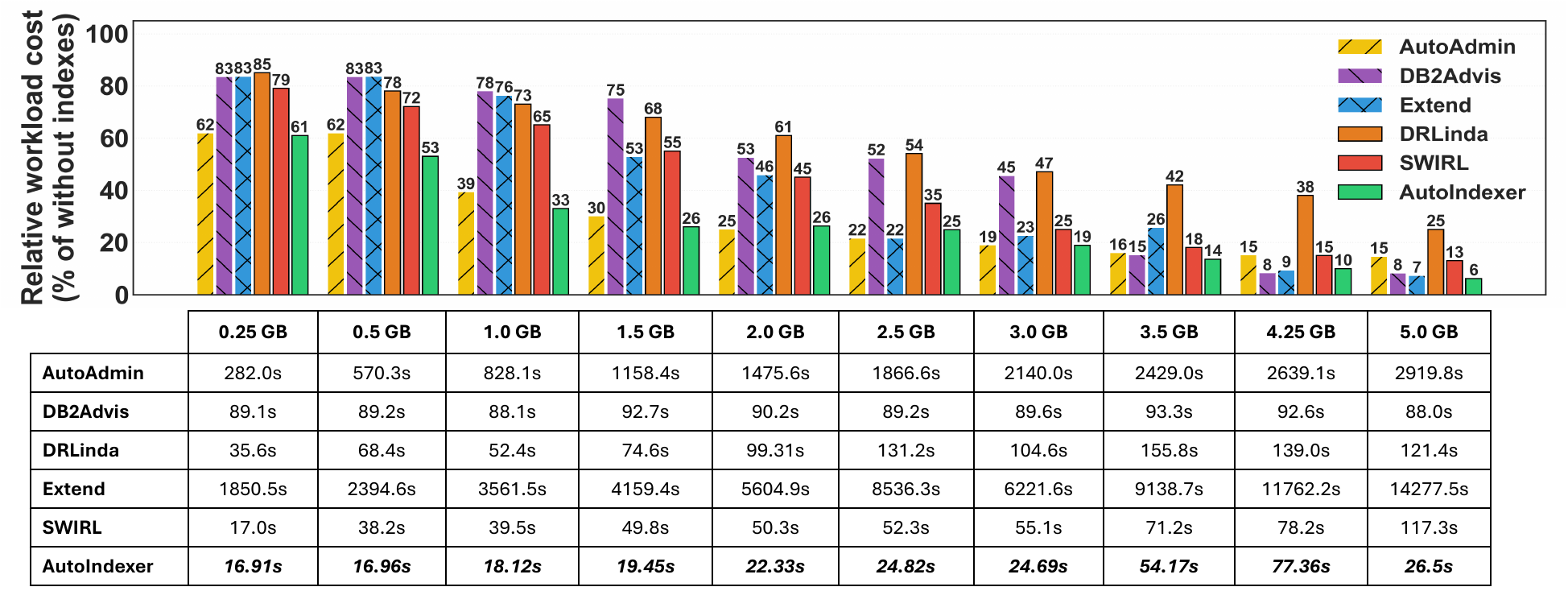}
\caption{Comparison of comparative approaches vs \emph{AutoIndexer} for a Scaling Join Order Benchmark workload on PostgreSQL. \textbf{Chart}: cost relative to processing w/o indexes; \textbf{Table}: selection runtime.}
\label{fig:perf_comparison}
\end{figure*}

\begin{figure}[htbp]
    \centering
    \includegraphics[width=1.0\linewidth]{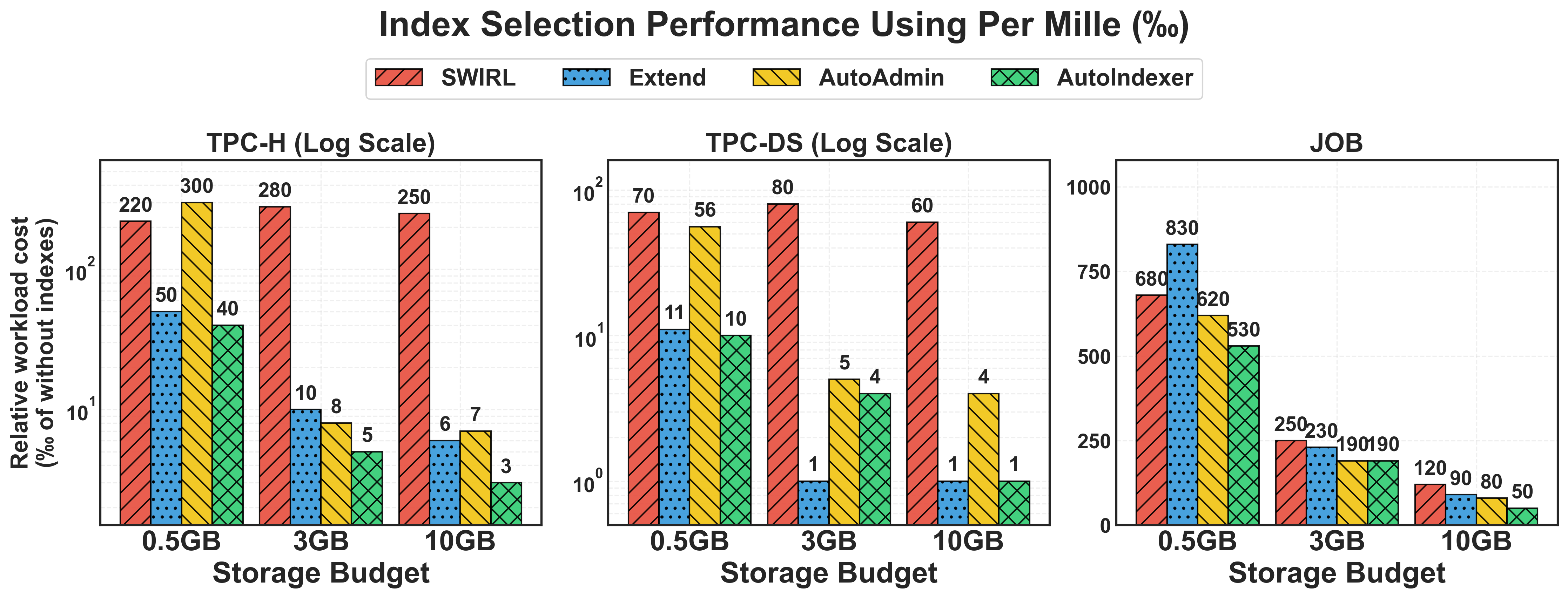}
    \caption{End2End workload runtime improvements among typical approaches across benchmarks}
    \label{fig:cross_bench}
\end{figure}

We performed comprehensive experiments to evaluate workload runtime cost improvements across multiple benchmarks, as shown in Figures~\ref{fig:perf_comparison} and~\ref{fig:cross_bench}, . Figure~\ref{fig:perf_comparison} focuses on a scaling Join Order Benchmark (JOB) under PostgreSQL, comparing relative workload cost (as a percentage of no-index baseline) and total index selection time for storage budgets between 0.25\,GB and 5\,GB. The classical approach \emph{Extend} requires over two hours at larger budgets due to a relatively more exhaustive search, while \emph{AutoAdmin} suffers substantial overhead. In contrast, \emph{AutoIndexer} achieves sub-minute selection times with near-optimal cost reductions. 

Figure~\ref{fig:cross_bench} extends these findings to TPC-H, TPC-DS, and additional JOB scenarios, highlighting how each benchmark’s structure affects both compressibility and RL-based indexing. In TPC-H, pruning redundant columns yields consistent gains—even at a modest 0.5\,GB budget, \emph{AutoIndexer} achieves 40‰ while \emph{AutoAdmin} remains at 300‰ (over 7× higher) and \emph{Extend} can only achieve 50\% relative costs compared with no-index cases. 
For TPC-DS, with its more complex joins and dimension tables, \emph{AutoIndexer} still provides roughly 5× improvements compared to baselines like \emph{SWIRL} (e.g., 10‰ vs 50–60‰ at certain budgets). 
Meanwhile, JOB’s real-world queries and correlated columns pose the greatest challenge, yet \emph{AutoIndexer} reduces final costs by half or more relative to competing methods. These cross-benchmark results confirm that our integrated pipeline adapts effectively across diverse schema complexities.

\subsubsection{Training efficiency and robustness}
\label{sec:training}

To evaluate training efficiency across varying workload sizes, we constructed training sets in JOB ranging from $N=80$ queries up to $N=1120$ and compared our indexer with other RL-based competitors. As shown in Table~\ref{tab:scaled_training}, the action space expands from 750 to 4813 when $W_{\max}=3$. Despite this substantial growth in index candidates, \emph{AutoIndexer} converged in just 2.1\,h at $N=1120$, whereas Lan et~al.\ and \emph{SWIRL} both struggled beyond $N=100$, often requiring 10--15+\,h or failing to converge at all.

Although \emph{SWIRL} shares a similar RL foundation, its exploration was too inefficient to scale. Even with double the training time of \emph{AutoIndexer}, \emph{SWIRL} frequently stalled or yielded suboptimal indexes for $N>100$, repeatedly re-testing partial index configurations and low-value actions, especially under correlated columns. In contrast, \emph{AutoIndexer}’s selective action pruning and compressed state representation targeted high-impact actions, enabling steady progress as workloads grew. This scalability is vital for real-world deployments, which demand rapid adaptation to shifting queries. To evaluate robustness, our training adheres to two key principles: (1) we withhold 20\% of query templates entirely from the training phase to ensure genuine generalization, and (2) although the workload scale matches the evaluation set, we generate a different set of queries from those templates to prevent overfitting. This approach ensures both template-level generalization and robustness to query variations. When randomly replacing 20\% to 50\%-80\% of seen templates/queries, performance drops by 13\%-21\% respectively, but still outperforms \emph{Extend} and \emph{AutoIndexer} by 30\% on solution quality (taking TPCH workload as an example). The selection time increases by 10\%-20\% but remains significantly lower than heuristic/traditional index advisors. This methodology validates both template-level generalization and robustness to diverse workload complexities and distributions.

While \emph{AutoIndexer} demonstrates strong performance, RL-based index selection inherently involves several trade-offs that warrant consideration in terms of training trade-offs and limitations. First, the initial training overhead is non-trivial: even with our compressed workloads, training may take hours versus immediate recommendations from traditional methods, making it prohibitive for rapidly changing schemas or highly dynamic workloads. Second, the approach assumes relatively stable query patterns during training, so entirely new query types can degrade performance until retraining occurs. Third, while compression techniques significantly reduce the action space, the RL agent still requires periodic retraining to adapt to substantial workload shifts, and concurrent query sessions during training can introduce noise in reward signals, affecting convergence stability. These limitations suggest that \emph{AutoIndexer} is most effective in environments with semi-stable workloads where the training investment can be amortized over extended periods, rather than highly volatile OLTP systems with rapid schema evolution.

\begin{table}[t]
\captionsetup{font=small}
\centering
\footnotesize
\caption{Training time comparison on JOB benchmark with increasing workload sizes till convergence ($W_{max}=3$), Storage Budget=10GB}
\label{tab:scaled_training}
\begin{tabular}{lrrrr}
\toprule
\#Queries & \#Actions & \multicolumn{3}{c}{Training Time (h)} \\
\cmidrule(lr){3-5}
& (A = I) & AutoIndexer & \makecell{Lan~\cite{lan2020index}} & \makecell{SWIRL~\cite{kossmann2022swirl}} \\
\midrule
80  & 750  & 0.3  & 1.5  & 5.9 \\
100 & 819  & 0.5  & 1.7  & 6.1 \\
500 & 2251 & 1.7  & 3.3  & 11.1 \\
1120& 4813 & 2.1 & 5.3   & >15 \\
\bottomrule
\end{tabular}
\end{table}


\subsubsection{Runtime Breakdown}
\label{sec:breakdown}

Table~\ref{tbl:runtime_breakdown} presents a selection runtime breakdown on the JOB workload under a 2\,GB budget, excluding any training overhead. Each system’s total time is partitioned into its core \emph{Algorithm} phase (including compression if applicable), \emph{Index} actions (e.g., physical index creation), and any remaining \emph{Other} overhead.

Overall, \emph{AutoIndexer} completes the entire process in under 25\,s, primarily due to its efficient compressor and specialized RL module, both of which help avoid superfluous exploration. In contrast, \emph{Extend}~\cite{schlosser2019efficient} reaches a much higher total time from an exhaustive enumeration that quickly becomes costly for large queries. Similarly, \emph{AutoAdmin}~\cite{chaudhuri1997efficient} requires substantial algorithmic time despite moderate index-creation overhead. Meanwhile, \emph{DB2Advis}~\cite{valentin2000db2} falls near the middle, trading some speed for simpler heuristics. Notably, \emph{SWIRL}~\cite{kossmann2022swirl} benefits from a reinforcement learning approach but remains slower than \emph{AutoIndexer}, underscoring the impact of \emph{AutoIndexer}’s compression and RL-model synergy. These comparisons confirm that our integrated design significantly cuts down selection time while maintaining or improving index quality.

\begin{table}[htbp]
\captionsetup{font=small}
\centering
\footnotesize
\caption{Selection runtime breakdown at a 2\,GB budget on JOB. Training time is excluded.}
\label{tbl:runtime_breakdown}
\begin{tabular}{@{}lcccc@{}}
\toprule
\textbf{System} & \textbf{\makecell{Algorithm(s) \\ (inc. Compress.)}} & \textbf{\makecell{Index \\ (s)}} & \textbf{\makecell{Other \\ (s)}} & \textbf{\makecell{Total \\ (s)}} \\
\midrule
AutoAdmin~\cite{chaudhuri1997efficient}   & 1440.0 & 20.4   & 15.2    & 1475.6 \\
DB2Advis~\cite{valentin2000db2}           & 79.0   & 7.1    & 4.2     & 90.3   \\
Extend~\cite{schlosser2019efficient}      & 5524.9 & 0.44   & 79.56   & 5604.9 \\
SWIRL~\cite{kossmann2022swirl}            & 44.14  & 0.20   & 5.96    & 50.3   \\
AutoIndexer                               & 16.91 (13.3)  & 0.15   & 5.27    & 22.33  \\
\bottomrule
\vspace{-10pt}
\end{tabular}
\end{table}

\subsubsection{Sole Effects AutoIndexer's Compression and RL model}
\label{sec:RL+Combine}

This section shows how we can \emph{mix and replace} individual \emph{AutoIndexer} components with different downstream index-selection algorithms or compression methods, thus isolating the contributions of our \textbf{(i) compressor} and \textbf{(ii) tailored State-wise RL model}.

\begin{figure}[htbp]
    \centering
    \includegraphics[width=0.94\linewidth]{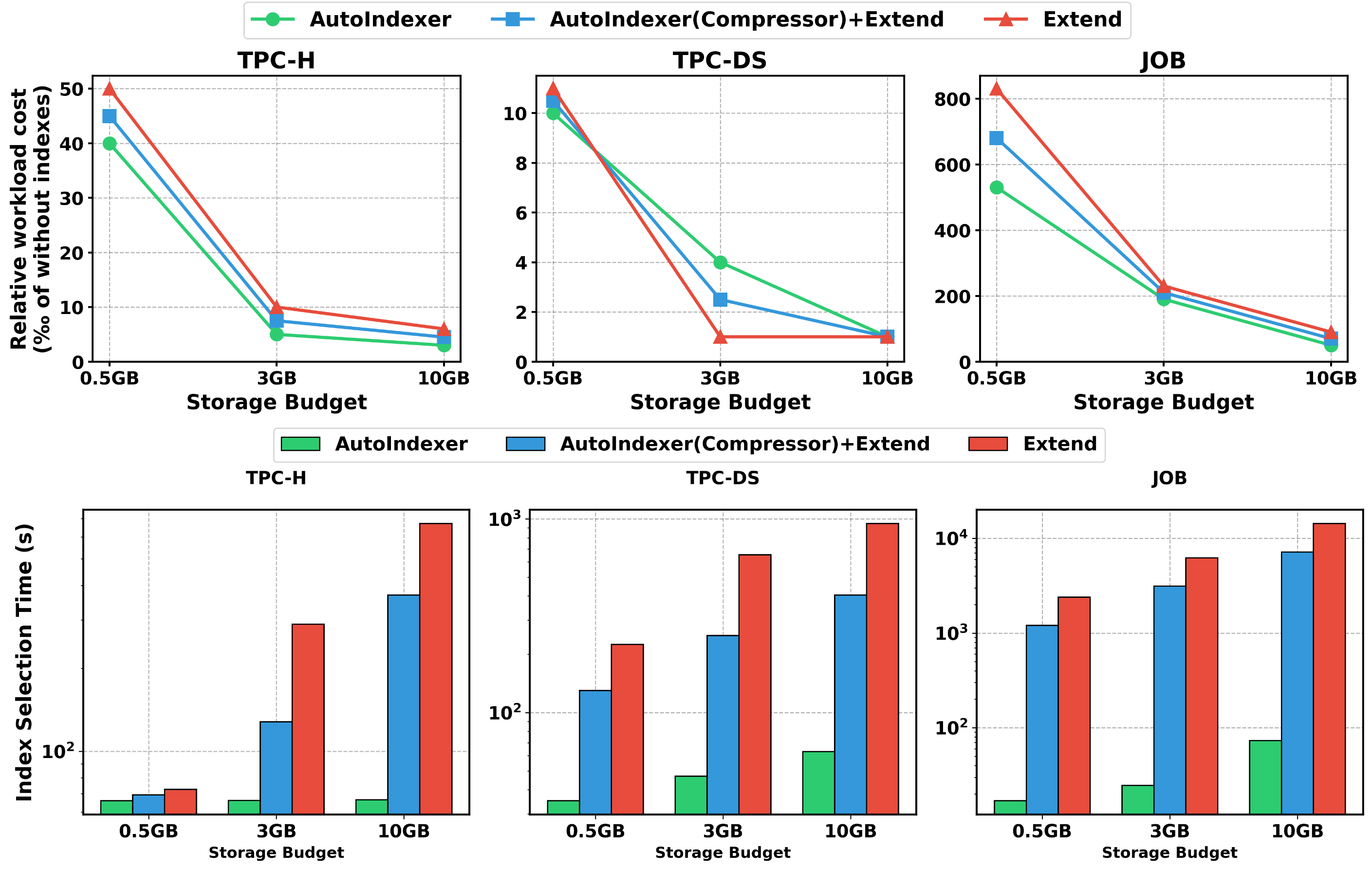}
    \caption{Sole effects on the compressor, combined with \emph{Extend}~\cite{schlosser2019efficient} and \emph{AutoIndexer} (Figure below is in Log Scale)}
    \label{fig:combine1}
    \vspace{-10pt}
\end{figure}

\begin{figure}[htbp]
    \centering
    \includegraphics[width=0.94\linewidth]{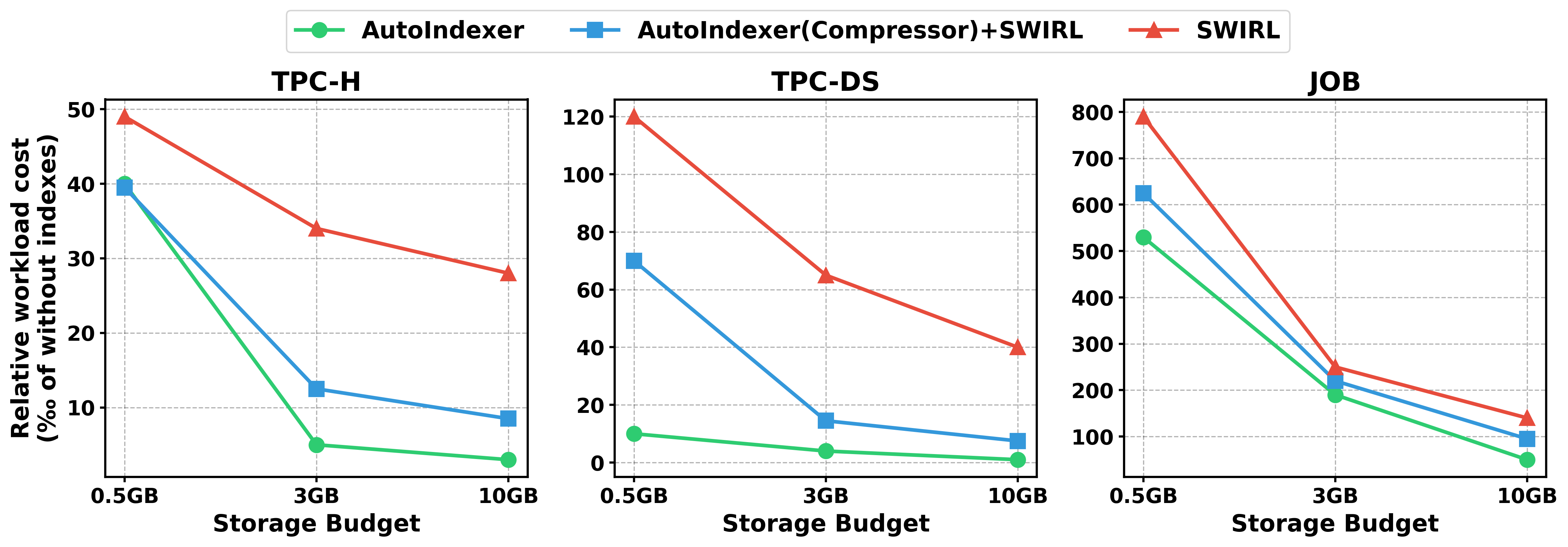}
    \caption{Sole effects on the compressor, combined with \emph{SWIRL}~\cite{kossmann2022swirl} and \emph{AutoIndexer}}
    \label{fig:combine2}
    \vspace{-12pt}
\end{figure}

\paragraph{(1) Sole Effects of the Compressor.}
To assess how our tailored compressor performs \emph{in isolation}, we conducted two sets of experiments:

\begin{itemize}[leftmargin=5pt]
    \item \textbf{Combining our compressor with other index-selection algorithms.} 
    Figure~\ref{fig:combine1} pairs the \emph{AutoIndexer} compressor with a classic heuristic–recursive strategy \emph{Extend}~\cite{schlosser2019efficient}, and Figure~\ref{fig:combine2} integrates the same compressor with \emph{SWIRL}~\cite{kossmann2022swirl}, a representative RL-based system. In both scenarios, we see that injecting our compressor yields considerable selection-time savings at smaller budgets, since naive enumeration becomes less overwhelming. However, the improvements are less dramatic than in \emph{AutoIndexer}’s native pipeline because our compressor is optimized for \emph{AutoIndexer}’s RL flow. Specifically, \emph{Extend} still faces a relatively large candidate space, and \emph{SWIRL} employs its own rule-based action pruning that does not fully exploit our compressor’s fine-grained query merges. Even so, we observe that compressing the workload helps both \emph{Extend} and \emph{SWIRL} moderate their overhead, proving that our compression is modular and can be ported across different selection algorithms.

    \item \textbf{Replacing our compressor with \textsc{Isum}’s}
    Beyond mixing our compressor into other algorithms, we also \emph{replaced} our module with \textsc{Isum}’s compression technique to highlight the difference in design philosophies. As summarized in Tables~\ref{tbl:comparison1} and~\ref{tbl:comparison2}, substituting \textsc{Isum}-style compression significantly increases both selection time and workload cost, often exceeding the slowest results under \emph{AutoIndexer}’s native compression. This gap remains large even when \textsc{Isum} compresses to different percentages, indicating that simply reducing workload size is not enough if the merges do not preserve key correlations. 
\end{itemize}

In short, these experiments verify our compressor’s standalone benefits. Although it can be mixed with heuristic or alternative RL-based pipelines for moderate gains, the synergy is stronger within \emph{AutoIndexer}’s complete workflow, where the compressor’s query merges are tailored to feed into our state-wise RL logic.

\begin{table}[ht]
\vspace{-5pt}
\captionsetup{font=small}
\centering
\footnotesize
\caption{Total Index Selection Time Comparison (s). ISUM-X indicates workload compression to \emph{X}\textperthousand \ of original size using \textsc{Isum} compression technique. Lower values indicate better performance.}
\vspace{-5pt}
\begin{tabular}{lcccc}
\toprule
\textbf{Workload} & \textbf{AutoIndexer} & \textbf{\makecell{ISUM\\-50}} & \textbf{\makecell{ISUM\\-100}} & \textbf{\makecell{ISUM\\-400}} \\
\midrule
JOB & 24.69 & 98.09  & 119.23  & 112.29  \\
TPC-DS & 47.01 & 220.31  & 84.24  & 117.33  \\
TPC-H & 66.08 & 131.78  & 133.97  & 136.41  \\
\bottomrule
\end{tabular}
\label{tbl:comparison1}
\end{table}

\begin{table}[ht]
\vspace{-20pt}
\captionsetup{font=small}
\centering
\footnotesize
\caption{Relative costs (in \%) compared to non-index situations.}
\begin{tabular}{lcccc}
\toprule
\textbf{Workload} & \textbf{AutoIndexer} & \textbf{\makecell{ISUM\\-50}} & \textbf{\makecell{ISUM\\-100}} & \textbf{\makecell{ISUM\\-400}} \\
\midrule
JOB & 19.0 & 39.4 & 33.7 & 33.0 \\
TPC-DS & 0.4 & 5.3 & 4.7 & 4.8 \\
TPC-H & 0.5 & 10.6 & 4.9 & 3.6 \\
\bottomrule
\vspace{-10pt}
\end{tabular}
\label{tbl:comparison2}
\end{table}

\begin{figure}[htbp]
\captionsetup{font=small}
    \centering
    \includegraphics[width=\linewidth]{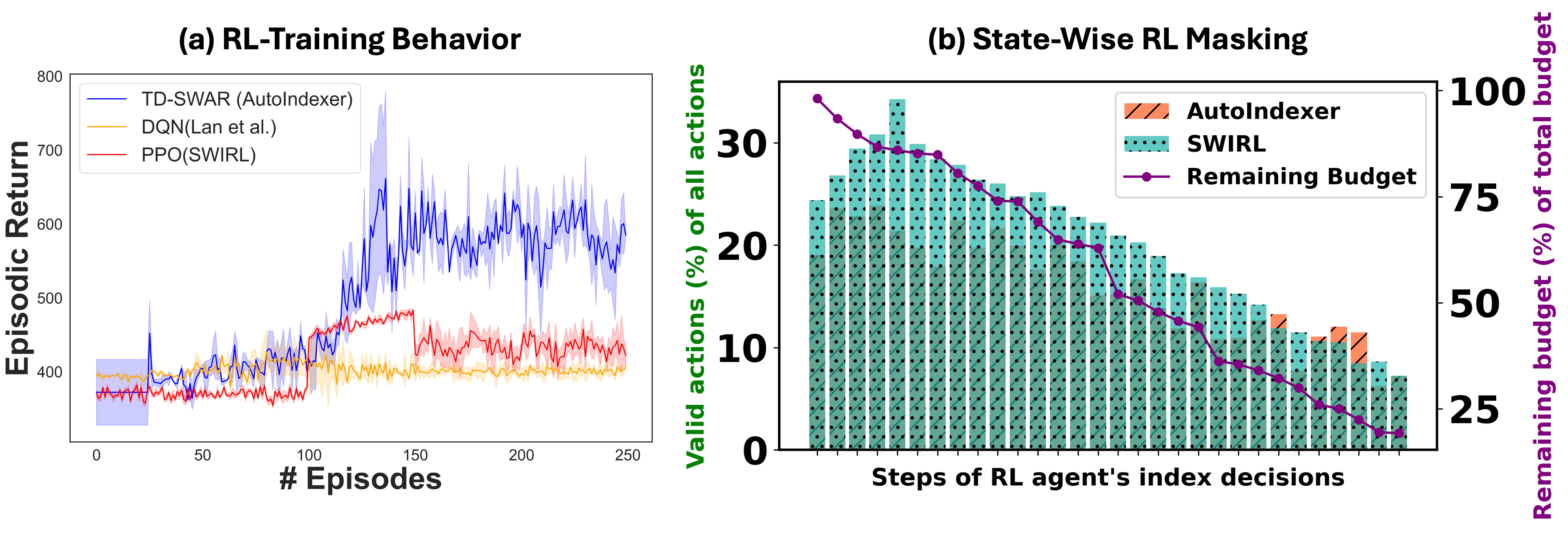}
    \vspace{-15pt}
    \caption{(a) Training behavior comparison between our proposed RL models and replacement RL models from prior works under identical settings. (b) Sole effects on the state-wise RL adaptive action masking with the same compressor on \emph{SWIRL}~\cite{kossmann2022swirl} and \emph{AutoIndexer}, maximum index width $W_{max} =5$}
    \label{fig:mask}
    \vspace{-10pt}
\end{figure}

\paragraph{(2) Isolating the RL Model’s Contributions.}
Having demonstrated how compression alone affects downstream algorithms, we now \emph{replace} the RL model within the \emph{AutoIndexer} pipeline (but \emph{keep} our compressor) to isolate the impact of our specialized RL design. Figure~\ref{fig:mask} highlights two aspects:

\begin{itemize}[leftmargin=5pt]
    \item \textbf{Training Behavior (Figure~\ref{fig:mask}\,(a)):} 
    We compare \emph{TD-SWAR (AutoIndexer)} with \emph{DQN}~\cite{van2016deep} (used by Lan et al.) and \emph{PPO}~\cite{schulman2017proximal} (the core of \emph{SWIRL}~\cite{kossmann2022swirl}). Early on, TD-SWAR and DQN appear similar, but TD-SWAR escalates more quickly, stabilizing at a higher episodic return beyond episode~100. PPO, by contrast, remains conservative and plateaus lower. This indicates that our RL model’s combination of target-based updates and state-wise indexing heuristics adapts more effectively, particularly once the compressed queries reduce repetitive exploration.

    \item \textbf{State-Wise RL Masking (Figure~\ref{fig:mask}\,(b)):}
    We then measure \emph{action pruning} behavior as storage budgets decrease to demonstrate how \emph{DEAS} is managed during deployment through our trained masking module. While \emph{SWIRL} (patterned teal bars) relies on rule-based heuristics to gradually reduce its valid action set, our RL agent (striped orange) employs more sophisticated logic that considers both compressed workload characteristics and remaining budget allocation. This targeted masking approach enables early elimination of suboptimal or infeasible indexes, leading to more efficient exploration. Although both approaches eventually achieve similar masking efficiency in later steps, \emph{AutoIndexer} identifies high-impact index subsets much earlier in the decision process, demonstrating more efficient action space pruning than \emph{SWIRL}.
\end{itemize}

These experiments show that, once the workload is compressed, our RL model in \emph{AutoIndexer} outperforms other RL cores by discarding irrelevant actions, finding stronger index configurations more quickly, and maintaining stable performance across different budgets.

\vspace{-12pt}

\subsubsection{Ablation studies}
\label{sec:ablation}

\begin{table*}[ht] 
\captionsetup{font=small}
\centering
\footnotesize
\caption{RL-based index advisors comparison}
\vspace{-10pt}
\label{tbl:comparison}
\begin{tabular}{@{}l|cccccc@{}}
 \toprule
 & DRLinda~\cite{sadri2020drlindex} & Lan~\cite{lan2020index} & SWIRL~\cite{kossmann2022swirl} & IA2~\cite{wang2024ia2} & DRLISA~\cite{yan2021index} & AutoIndexer \\
 \midrule
 Multi-attribute & No & Yes & Yes & Yes & Yes & Yes \\
 Stop criterion & \#Idx & \#Idx,Storage & Storage & Storage & No-Improvement.& Storage \\
 Workload Modeling & Yes & No & Yes & Yes & No & Yes \\
 Generalization & ++ & - & +++ & +++ & - & +++ \\
 Training & ++ & + & +++ & ++ & Unspecified & + \\
 Scaling Awareness & N/A & Rule Mask. & Rule Mask. & Adapt. Mask. & N/A & Auto Compression + Adapt. Mask \\
 \bottomrule
\end{tabular}
\vspace{-5pt}
\end{table*}

\begin{figure}[htbp]
    \centering
    \includegraphics[width=0.93\linewidth]{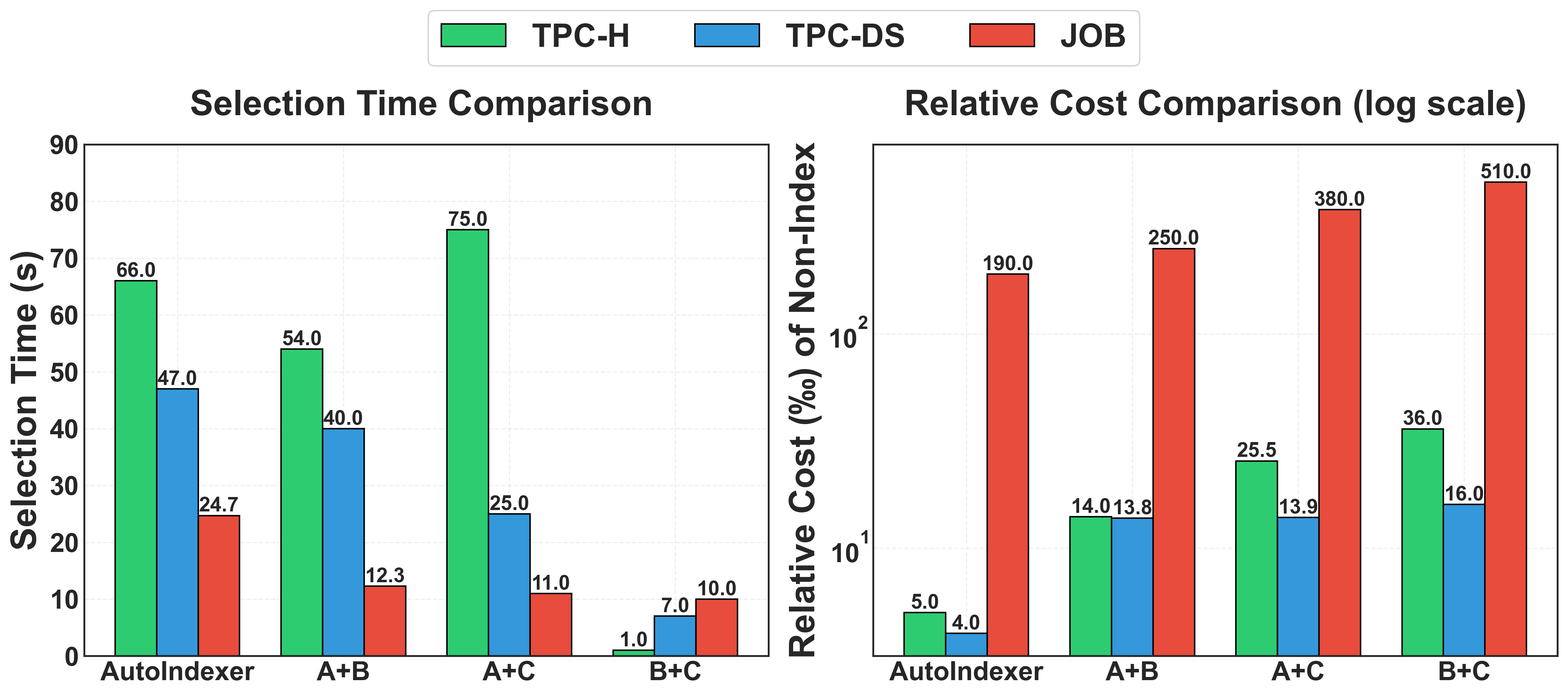}
    \caption{Ablation Study: Component Impact Analysis (3GB Budget). \textbf{(A)-compressor} , \textbf{(B)-RL model}, and \textbf{(C)-workload model}.}
    \label{fig:ablation}
    \vspace{-5pt}
\end{figure}

Figure~\ref{fig:ablation} compares four \emph{AutoIndexer} variants by omitting or replacing its core modules: \textbf{(A)} the column-centric compressor, \textbf{(B)} our specialized RL (TD-SWAR), and \textbf{(C)} the workload model. Removing \textbf{C} (A+B) forces simple tokenized queries instead of learned statistics, causing a 9\% performance drop for JOB while simpler benchmarks show moderate impact. This gap reveals how workload modeling becomes critical for capturing complex query relationships and column correlations in real-world scenarios.

Replacing \textbf{B} (A+C) with grid-search+greedy approach increases TPC-H's cost from 5‰ to 25‰, and similarly degrades performance on other benchmarks. This demonstrates that specialized RL is essential for navigating the complex trade-offs between index utility and storage costs, particularly when dealing with multi-attribute indexes. Most notably, discarding the compressor (\textbf{B+C}) shows a deceptive trade-off: while selection time decreases, query costs soar significantly (JOB by 2.7×, TPC-H by 7×, TPC-DS by 4×). This severe performance degradation stems from the RL model struggling with an expanded, less-structured search space that hampers effective exploration and convergence. Overall, each module clearly addresses a distinct bottleneck: \textbf{(A)} shrinks the candidate space, \textbf{(B)} optimizes cost-effective indexes under repetitive trials, and \textbf{(C)} injects schema-level insights crucial for complex workloads like JOB. Removing any one of them yields noticeable regressions in both overhead and query performance.

\section{Related Works}
\label{sec:related}

\subsection{RL for Broader Database Tuning}
\label{sec:relwork-rl-other}

Reinforcement learning now permeates many aspects of physical-design and run-time optimisation for relational DBMSs.  For query planning, deep-RL join optimisers—DQO~\cite{Marcus_aiDM18}, Tree-LSTM ordering~\cite{Yu_ICDE20}, graph policies~\cite{Chen_KDD22}, and the regret-bounded \emph{SkinnerDB} engine~\cite{Trummer_SIGMOD19}—routinely outperform classical dynamic programming on complex join graphs.  In physical design, RL agents advise materialised views: \emph{AutoView}~\cite{Han_ICDE21}, opportunistic view materialisation~\cite{Liang_arXiv19}, and joint view–plus–index selection~\cite{Wang_ICEBE24} balance maintenance overhead against query speed-up.  System-knob tuning has likewise embraced ML/RL pipelines, from \emph{OtterTune}’s large-scale models~\cite{VanAken_SIGMOD17} to cloud-native RL optimisers~\cite{Zhang_SIGMOD19} and the query-aware \emph{QTune}~\cite{Li_PVLDB19}.  Learned query optimisers—including \emph{Neo}~\cite{Marcus_PVLDB19}, \emph{Bao}~\cite{Marcus_SIGMOD21}, \emph{Balsa}~\cite{Yang_SIGMOD22}, and \emph{Lero}~\cite{Zhu_PVLDB23}—replace hand-crafted cost models with neural policies, while hybrids merge ML with classic plan search~\cite{Yu_PVLDB22}.  These advances build on foundations laid by IBM’s \emph{LEO}~\cite{Stillger_VLDB01} and are surveyed in recent overviews of learned self-management~\cite{sharma2018case}.

\vspace{-5pt}
\subsection{Traditional Index Selection Approaches}
\label{sec:relwork-traditional}

Index selection has been studied for over five decades~\cite{lum1971optimization}, yielding a wide range of \emph{reductive} and \emph{additive} strategies~\cite{bruno2005automatic,whang1987index,chaudhuri1997efficient,valentin2000db2,schlosser2019efficient}. 
\emph{Reductive} methods, such as the approaches by Bruno et al.~\cite{bruno2005automatic} and Whang et al.~\cite{whang1987index}, begin with a comprehensive set of candidate indexes and iteratively prune those deemed less valuable. 
In contrast, \emph{additive} techniques like \emph{AutoAdmin}~\cite{chaudhuri1997efficient} and \emph{DB2Advis}~\cite{valentin2000db2} build up from an empty set, adding promising indexes until storage or performance budgets are reached, while \emph{Extend}~\cite{schlosser2019efficient} refines cost estimates with heuristic adjustments to find near-optimal solutions in practice. Although these algorithms often excel on moderate workloads~\cite{valentin2000db2,schlosser2019efficient}, they can falter for large-scale or complex query mixes~\cite{cooper2010benchmarking}, as no single heuristic simultaneously achieves top-tier solution quality and fast enumeration. 
\emph{Adaptive indexing}~\cite{idreos2011merging} brings incremental online updates to certain column-store systems but is not readily compatible with hypothetical index frameworks in most relational DBMSs~\cite{HypoPG,brucato2024wred,siddiqui2022isum}. Consequently, scaling concerns, heterogeneous query structures, and complex index interactions remain open challenges for these long-standing methods.

\vspace{-5pt}
\subsection{RL-based Index Selection Approaches}
\label{sec:relwork-rl}

Recent RL-based index advisors have shown encouraging results, yet each carries limitations that hinder practical adoption (Table~\ref{tbl:comparison}).  
\emph{DRLinda}~\cite{sadri2020drlindex} employs multi-agent RL but supports only single-attribute indexes and lacks a public implementation. Lan\,\textit{et al.}~\cite{lan2020index} extend support to multi-attribute indexes, though at the cost of a heavily pruned candidate pool that impairs generalization to unseen queries.   \emph{DRLISA}~\cite{yan2021index}, aimed at NoSQL workloads, terminates once no further improvement is observed and provides few details on its workload encoding.  
More sophisticated solutions such as \emph{SWIRL}~\cite{kossmann2022swirl} and \emph{IA2}~\cite{wang2024ia2} integrate action masking and richer workload embeddings, yet incur substantial training overhead and scale poorly to large, dynamic OLAP workloads.  
In all cases, the \emph{Disaster of Exploded Action Space} (DEAS)~\cite{andriotis2019managing,majeed2021exact} remains an open bottleneck, especially when multi-attribute indexes are considered.

\vspace{-5pt}
\subsection{Workload Compression}
\label{sec:workload-compression}

Recent techniques have demonstrated how workload compression can reduce index tuning overhead. \textsc{Gsum}~\cite{deep2020comprehensive} introduced an early-stage indexing-agnostic approach, while its successor \textsc{Isum}~\cite{siddiqui2022isum} identifies representative query subsets through lightweight scoring mechanisms. Although \textsc{Isum} typically produces high-quality solutions, its frequent reliance on cost models and \emph{"what-if"} analyses tends to slow the tuning process.

\textsc{WRED}~\cite{brucato2024wred} takes a different approach by employing statistic-based modeling without optimizer involvement. By analyzing the abstract syntax tree representation of SQL queries, the method relies on heuristic-based query-level reductions involving complex structure and AST traversals to achieve workload compression. While \textsc{WRED} enables quick compression at the cost of diminished solution quality—particularly when paired with suboptimal downstream tuners—its purely heuristic methodology focuses on query-level operations rather than column-oriented strategies.

Although both approaches accelerate traditional cost-based advisors, they depend heavily on static heuristics and user-defined parameters. More importantly, neither \textsc{Isum} nor \textsc{WRED} considered RL-based index selection within their designs. Large action spaces and dynamic reward structures in RL-based systems demand compression strategies that go beyond fixed heuristics and pairwise comparisons.


\vspace{-5pt}
\section{Conclusion}
\label{sec:conclusion}

In this paper, we presented \emph{AutoIndexer}, a novel framework that combines intelligent workload compression with specialized RL models to tackle index selection at scale. \emph{AutoIndexer} shows that the twin problems that have long plagued reinforcement learning-based index advisers—oversized workloads and the disaster of exploded action spaces—can be solved \emph{jointly}.

The framework is modular: the graph compressor could be extended to materialized views, the selector network can absorb richer priors, and the vectorized env is ready for continual-learning scenarios.  Taken together, these results mark a practical step toward fully autonomous physical-design tuning in large-scale data systems.

\section*{Acknowledgments}

We thank Zak Singh for his invaluable assistance with proofreading and manuscript preparation. His careful review and feedback significantly improved the clarity and presentation of this work.

\bibliographystyle{ACM-Reference-Format}
\bibliography{sample}


\begin{thebibliography}{51}


\ifx \showCODEN    \undefined \def \showCODEN     #1{\unskip}     \fi
\ifx \showDOI      \undefined \def \showDOI       #1{#1}\fi
\ifx \showISBNx    \undefined \def \showISBNx     #1{\unskip}     \fi
\ifx \showISBNxiii \undefined \def \showISBNxiii  #1{\unskip}     \fi
\ifx \showISSN     \undefined \def \showISSN      #1{\unskip}     \fi
\ifx \showLCCN     \undefined \def \showLCCN      #1{\unskip}     \fi
\ifx \shownote     \undefined \def \shownote      #1{#1}          \fi
\ifx \showarticletitle \undefined \def \showarticletitle #1{#1}   \fi
\ifx \showURL      \undefined \def \showURL       {\relax}        \fi
\providecommand\bibfield[2]{#2}
\providecommand\bibinfo[2]{#2}
\providecommand\natexlab[1]{#1}
\providecommand\showeprint[2][]{arXiv:#2}

\bibitem[Aken et~al\mbox{.}(2017)]%
        {VanAken_SIGMOD17}
\bibfield{author}{\bibinfo{person}{Dana~Van Aken}, \bibinfo{person}{Andrew Pavlo}, \bibinfo{person}{Geoffrey~J. Gordon}, {and} \bibinfo{person}{Bohan Zhang}.} \bibinfo{year}{2017}\natexlab{}.
\newblock \showarticletitle{Automatic Database Management System Tuning Through Large-Scale Machine Learning}. In \bibinfo{booktitle}{\emph{Proceedings of the 2017 ACM SIGMOD International Conference on Management of Data}}. \bibinfo{pages}{1009--1024}.
\newblock


\bibitem[Andriotis and Papakonstantinou(2019)]%
        {andriotis2019managing}
\bibfield{author}{\bibinfo{person}{Charalampos~P Andriotis} {and} \bibinfo{person}{Konstantinos~G Papakonstantinou}.} \bibinfo{year}{2019}\natexlab{}.
\newblock \showarticletitle{Managing engineering systems with large state and action spaces through deep reinforcement learning}.
\newblock \bibinfo{journal}{\emph{Reliability Engineering \& System Safety}}  \bibinfo{volume}{191} (\bibinfo{year}{2019}), \bibinfo{pages}{106483}.
\newblock


\bibitem[Brucato et~al\mbox{.}(2024)]%
        {brucato2024wred}
\bibfield{author}{\bibinfo{person}{Matteo Brucato}, \bibinfo{person}{Tarique Siddiqui}, \bibinfo{person}{Wentao Wu}, \bibinfo{person}{Vivek Narasayya}, {and} \bibinfo{person}{Surajit Chaudhuri}.} \bibinfo{year}{2024}\natexlab{}.
\newblock \showarticletitle{Wred: Workload Reduction for Scalable Index Tuning}.
\newblock \bibinfo{journal}{\emph{Proceedings of the ACM on Management of Data}} \bibinfo{volume}{2}, \bibinfo{number}{1} (\bibinfo{year}{2024}), \bibinfo{pages}{1--26}.
\newblock


\bibitem[Bruno and Chaudhuri(2005)]%
        {bruno2005automatic}
\bibfield{author}{\bibinfo{person}{Nicolas Bruno} {and} \bibinfo{person}{Surajit Chaudhuri}.} \bibinfo{year}{2005}\natexlab{}.
\newblock \showarticletitle{Automatic physical database tuning: A relaxation-based approach}. In \bibinfo{booktitle}{\emph{Proceedings of the 2005 ACM SIGMOD international conference on Management of data}}. \bibinfo{pages}{227--238}.
\newblock


\bibitem[Chaudhuri and Narasayya(1997)]%
        {chaudhuri1997efficient}
\bibfield{author}{\bibinfo{person}{Surajit Chaudhuri} {and} \bibinfo{person}{Vivek~R Narasayya}.} \bibinfo{year}{1997}\natexlab{}.
\newblock \showarticletitle{An efficient, cost-driven index selection tool for Microsoft SQL server}. In \bibinfo{booktitle}{\emph{VLDB}}, Vol.~\bibinfo{volume}{97}. San Francisco, \bibinfo{pages}{146--155}.
\newblock


\bibitem[Chen et~al\mbox{.}(2022)]%
        {Chen_KDD22}
\bibfield{author}{\bibinfo{person}{Jin Chen}, \bibinfo{person}{Guanyu Ye}, \bibinfo{person}{Yan Zhao}, \bibinfo{person}{Shuncheng Liu}, \bibinfo{person}{Liwei Deng}, \bibinfo{person}{Xu Chen}, \bibinfo{person}{Rui Zhou}, {and} \bibinfo{person}{Kai Zheng}.} \bibinfo{year}{2022}\natexlab{}.
\newblock \showarticletitle{Efficient Join Order Selection Learning with Graph-based Representation}. In \bibinfo{booktitle}{\emph{Proceedings of the 28th ACM SIGKDD International Conference on Knowledge Discovery \& Data Mining (KDD)}}. \bibinfo{pages}{97--107}.
\newblock


\bibitem[Chu et~al\mbox{.}(2018)]%
        {chu2018axiomatic}
\bibfield{author}{\bibinfo{person}{Shumo Chu}, \bibinfo{person}{Brendan Murphy}, \bibinfo{person}{Jared Roesch}, \bibinfo{person}{Alvin Cheung}, {and} \bibinfo{person}{Dan Suciu}.} \bibinfo{year}{2018}\natexlab{}.
\newblock \showarticletitle{Axiomatic foundations and algorithms for deciding semantic equivalences of SQL queries}.
\newblock \bibinfo{journal}{\emph{arXiv preprint arXiv:1802.02229}} (\bibinfo{year}{2018}).
\newblock


\bibitem[Contributors(2024)]%
        {HypoPG}
\bibfield{author}{\bibinfo{person}{HypoPG Contributors}.} \bibinfo{year}{2024}\natexlab{}.
\newblock \bibinfo{title}{HypoPG: Hypothetical Indexes for PostgreSQL}.
\newblock \bibinfo{howpublished}{\url{https://github.com/HypoPG/hypopg}}.
\newblock
\newblock
\shownote{Accessed: 2024-01-01}.


\bibitem[Cooper et~al\mbox{.}(2010)]%
        {cooper2010benchmarking}
\bibfield{author}{\bibinfo{person}{Brian~F Cooper}, \bibinfo{person}{Adam Silberstein}, \bibinfo{person}{Erwin Tam}, \bibinfo{person}{Raghu Ramakrishnan}, {and} \bibinfo{person}{Russell Sears}.} \bibinfo{year}{2010}\natexlab{}.
\newblock \showarticletitle{Benchmarking cloud serving systems with YCSB}. In \bibinfo{booktitle}{\emph{Proceedings of the 1st ACM symposium on Cloud computing}}. \bibinfo{pages}{143--154}.
\newblock


\bibitem[De~Moura and Bj{\o}rner(2008)]%
        {de2008z3}
\bibfield{author}{\bibinfo{person}{Leonardo De~Moura} {and} \bibinfo{person}{Nikolaj Bj{\o}rner}.} \bibinfo{year}{2008}\natexlab{}.
\newblock \showarticletitle{Z3: An efficient SMT solver}. In \bibinfo{booktitle}{\emph{International conference on Tools and Algorithms for the Construction and Analysis of Systems}}. Springer, \bibinfo{pages}{337--340}.
\newblock


\bibitem[Deep et~al\mbox{.}(2020)]%
        {deep2020comprehensive}
\bibfield{author}{\bibinfo{person}{Shaleen Deep}, \bibinfo{person}{Anja Gruenheid}, \bibinfo{person}{Paraschos Koutris}, \bibinfo{person}{Jeffrey Naughton}, {and} \bibinfo{person}{Stratis Viglas}.} \bibinfo{year}{2020}\natexlab{}.
\newblock \showarticletitle{Comprehensive and efficient workload compression}.
\newblock \bibinfo{journal}{\emph{arXiv preprint arXiv:2011.05549}} (\bibinfo{year}{2020}).
\newblock


\bibitem[Han et~al\mbox{.}(2021)]%
        {Han_ICDE21}
\bibfield{author}{\bibinfo{person}{Yue Han}, \bibinfo{person}{Guoliang Li}, \bibinfo{person}{Haitao Yuan}, {and} \bibinfo{person}{Ji Sun}.} \bibinfo{year}{2021}\natexlab{}.
\newblock \showarticletitle{AutoView: An Autonomous Materialized View Management System with Deep Reinforcement Learning}. In \bibinfo{booktitle}{\emph{Proceedings of the 37th IEEE International Conference on Data Engineering (ICDE)}}. \bibinfo{pages}{289--300}.
\newblock


\bibitem[Idreos et~al\mbox{.}(2009)]%
        {idreos2009self}
\bibfield{author}{\bibinfo{person}{Stratos Idreos}, \bibinfo{person}{Martin~L Kersten}, {and} \bibinfo{person}{Stefan Manegold}.} \bibinfo{year}{2009}\natexlab{}.
\newblock \showarticletitle{Self-organizing tuple reconstruction in column-stores}. In \bibinfo{booktitle}{\emph{Proceedings of the 2009 ACM SIGMOD International Conference on Management of data}}. \bibinfo{pages}{297--308}.
\newblock


\bibitem[Idreos et~al\mbox{.}(2011)]%
        {idreos2011merging}
\bibfield{author}{\bibinfo{person}{Stratos Idreos}, \bibinfo{person}{Stefan Manegold}, \bibinfo{person}{Harumi Kuno}, {and} \bibinfo{person}{Goetz Graefe}.} \bibinfo{year}{2011}\natexlab{}.
\newblock \showarticletitle{Merging what's cracked, cracking what's merged: adaptive indexing in main-memory column-stores}.
\newblock \bibinfo{journal}{\emph{Proceedings of the VLDB Endowment}} \bibinfo{volume}{4}, \bibinfo{number}{9} (\bibinfo{year}{2011}), \bibinfo{pages}{586--597}.
\newblock


\bibitem[Kossmann et~al\mbox{.}(2022)]%
        {kossmann2022swirl}
\bibfield{author}{\bibinfo{person}{Jan Kossmann}, \bibinfo{person}{Alexander Kastius}, {and} \bibinfo{person}{Rainer Schlosser}.} \bibinfo{year}{2022}\natexlab{}.
\newblock \showarticletitle{SWIRL: Selection of Workload-aware Indexes using Reinforcement Learning.}. In \bibinfo{booktitle}{\emph{EDBT}}, Vol.~\bibinfo{volume}{2}. \bibinfo{pages}{155--2}.
\newblock


\bibitem[Lan et~al\mbox{.}(2020)]%
        {lan2020index}
\bibfield{author}{\bibinfo{person}{Hai Lan}, \bibinfo{person}{Zhifeng Bao}, {and} \bibinfo{person}{Yuwei Peng}.} \bibinfo{year}{2020}\natexlab{}.
\newblock \showarticletitle{An index advisor using deep reinforcement learning}. In \bibinfo{booktitle}{\emph{Proceedings of the 29th ACM International Conference on Information \& Knowledge Management}}. \bibinfo{pages}{2105--2108}.
\newblock


\bibitem[Li et~al\mbox{.}(2019)]%
        {Li_PVLDB19}
\bibfield{author}{\bibinfo{person}{Guoliang Li}, \bibinfo{person}{Xuanhe Zhou}, \bibinfo{person}{Shifu Li}, {and} \bibinfo{person}{Bo Gao}.} \bibinfo{year}{2019}\natexlab{}.
\newblock \showarticletitle{QTune: A Query-Aware Database Tuning System with Deep Reinforcement Learning}.
\newblock \bibinfo{journal}{\emph{Proc. VLDB Endowment}} \bibinfo{volume}{12}, \bibinfo{number}{12} (\bibinfo{year}{2019}), \bibinfo{pages}{2118--2130}.
\newblock


\bibitem[Liang et~al\mbox{.}(2019)]%
        {Liang_arXiv19}
\bibfield{author}{\bibinfo{person}{Xi Liang}, \bibinfo{person}{Aaron~J. Elmore}, {and} \bibinfo{person}{Sanjay Krishnan}.} \bibinfo{year}{2019}\natexlab{}.
\newblock \showarticletitle{Opportunistic View Materialization with Deep Reinforcement Learning}.
\newblock \bibinfo{journal}{\emph{CoRR}}  \bibinfo{volume}{abs/1903.01363} (\bibinfo{year}{2019}).
\newblock


\bibitem[Lum and Ling(1971)]%
        {lum1971optimization}
\bibfield{author}{\bibinfo{person}{Vincent~Y Lum} {and} \bibinfo{person}{Huei Ling}.} \bibinfo{year}{1971}\natexlab{}.
\newblock \showarticletitle{An optimization problem on the selection of secondary keys}. In \bibinfo{booktitle}{\emph{Proceedings of the 1971 26th annual conference}}. \bibinfo{pages}{349--356}.
\newblock


\bibitem[Majeed and Hutter(2021)]%
        {majeed2021exact}
\bibfield{author}{\bibinfo{person}{Sultan~J Majeed} {and} \bibinfo{person}{Marcus Hutter}.} \bibinfo{year}{2021}\natexlab{}.
\newblock \showarticletitle{Exact reduction of huge action spaces in general reinforcement learning}. In \bibinfo{booktitle}{\emph{Proceedings of the AAAI Conference on Artificial Intelligence}}, Vol.~\bibinfo{volume}{35}. \bibinfo{pages}{8874--8883}.
\newblock


\bibitem[Marcus et~al\mbox{.}(2021)]%
        {Marcus_SIGMOD21}
\bibfield{author}{\bibinfo{person}{Ryan Marcus}, \bibinfo{person}{Parimarjan Negi}, \bibinfo{person}{Hongzi Mao}, \bibinfo{person}{Nesime Tatbul}, \bibinfo{person}{Mohammad Alizadeh}, {and} \bibinfo{person}{Tim Kraska}.} \bibinfo{year}{2021}\natexlab{}.
\newblock \showarticletitle{Bao: Making Learned Query Optimization Practical}. In \bibinfo{booktitle}{\emph{Proceedings of the 2021 ACM SIGMOD International Conference on Management of Data}}. \bibinfo{pages}{1275--1288}.
\newblock


\bibitem[Marcus et~al\mbox{.}(2019)]%
        {Marcus_PVLDB19}
\bibfield{author}{\bibinfo{person}{Ryan Marcus}, \bibinfo{person}{Parimarjan Negi}, \bibinfo{person}{Hongzi Mao}, \bibinfo{person}{Chi Zhang}, \bibinfo{person}{Mohammad Alizadeh}, \bibinfo{person}{Tim Kraska}, \bibinfo{person}{Olga Papaemmanouil}, {and} \bibinfo{person}{Nesime Tatbul}.} \bibinfo{year}{2019}\natexlab{}.
\newblock \showarticletitle{Neo: A Learned Query Optimizer}.
\newblock \bibinfo{journal}{\emph{Proc. VLDB Endowment}} \bibinfo{volume}{12}, \bibinfo{number}{11} (\bibinfo{year}{2019}), \bibinfo{pages}{1705--1718}.
\newblock


\bibitem[Marcus and Papaemmanouil(2018)]%
        {Marcus_aiDM18}
\bibfield{author}{\bibinfo{person}{Ryan Marcus} {and} \bibinfo{person}{Olga Papaemmanouil}.} \bibinfo{year}{2018}\natexlab{}.
\newblock \showarticletitle{Deep Reinforcement Learning for Join Order Enumeration}. In \bibinfo{booktitle}{\emph{Proceedings of the 1st Intl. Workshop on Exploiting AI Techniques for Data Management (aiDM@SIGMOD)}}. \bibinfo{pages}{1--4}.
\newblock


\bibitem[Niwattanakul et~al\mbox{.}(2013)]%
        {niwattanakul2013using}
\bibfield{author}{\bibinfo{person}{Suphakit Niwattanakul}, \bibinfo{person}{Jatsada Singthongchai}, \bibinfo{person}{Ekkachai Naenudorn}, {and} \bibinfo{person}{Supachanun Wanapu}.} \bibinfo{year}{2013}\natexlab{}.
\newblock \showarticletitle{Using of Jaccard coefficient for keywords similarity}. In \bibinfo{booktitle}{\emph{Proceedings of the international multiconference of engineers and computer scientists}}, Vol.~\bibinfo{volume}{1}. \bibinfo{pages}{380--384}.
\newblock


\bibitem[Russo et~al\mbox{.}(2018)]%
        {russo2018tutorial}
\bibfield{author}{\bibinfo{person}{Daniel~J Russo}, \bibinfo{person}{Benjamin Van~Roy}, \bibinfo{person}{Abbas Kazerouni}, \bibinfo{person}{Ian Osband}, \bibinfo{person}{Zheng Wen}, {et~al\mbox{.}}} \bibinfo{year}{2018}\natexlab{}.
\newblock \showarticletitle{A tutorial on thompson sampling}.
\newblock \bibinfo{journal}{\emph{Foundations and Trends{\textregistered} in Machine Learning}} \bibinfo{volume}{11}, \bibinfo{number}{1} (\bibinfo{year}{2018}), \bibinfo{pages}{1--96}.
\newblock


\bibitem[Sadri et~al\mbox{.}(2020)]%
        {sadri2020drlindex}
\bibfield{author}{\bibinfo{person}{Zahra Sadri}, \bibinfo{person}{Le Gruenwald}, {and} \bibinfo{person}{Eleazar Lead}.} \bibinfo{year}{2020}\natexlab{}.
\newblock \showarticletitle{DRLindex: deep reinforcement learning index advisor for a cluster database}. In \bibinfo{booktitle}{\emph{Proceedings of the 24th Symposium on International Database Engineering \& Applications}}. \bibinfo{pages}{1--8}.
\newblock


\bibitem[Schlosser et~al\mbox{.}(2019)]%
        {schlosser2019efficient}
\bibfield{author}{\bibinfo{person}{Rainer Schlosser}, \bibinfo{person}{Jan Kossmann}, {and} \bibinfo{person}{Martin Boissier}.} \bibinfo{year}{2019}\natexlab{}.
\newblock \showarticletitle{Efficient scalable multi-attribute index selection using recursive strategies}. In \bibinfo{booktitle}{\emph{2019 IEEE 35th International Conference on Data Engineering (ICDE)}}. IEEE, \bibinfo{pages}{1238--1249}.
\newblock


\bibitem[Schulman et~al\mbox{.}(2017)]%
        {schulman2017proximal}
\bibfield{author}{\bibinfo{person}{John Schulman}, \bibinfo{person}{Filip Wolski}, \bibinfo{person}{Prafulla Dhariwal}, \bibinfo{person}{Alec Radford}, {and} \bibinfo{person}{Oleg Klimov}.} \bibinfo{year}{2017}\natexlab{}.
\newblock \showarticletitle{Proximal policy optimization algorithms}.
\newblock \bibinfo{journal}{\emph{arXiv preprint arXiv:1707.06347}} (\bibinfo{year}{2017}).
\newblock


\bibitem[Sharma et~al\mbox{.}(2018)]%
        {sharma2018case}
\bibfield{author}{\bibinfo{person}{Ankur Sharma}, \bibinfo{person}{Felix~Martin Schuhknecht}, {and} \bibinfo{person}{Jens Dittrich}.} \bibinfo{year}{2018}\natexlab{}.
\newblock \showarticletitle{The case for automatic database administration using deep reinforcement learning}.
\newblock \bibinfo{journal}{\emph{arXiv preprint arXiv:1801.05643}} (\bibinfo{year}{2018}).
\newblock


\bibitem[Siddiqui et~al\mbox{.}(2022)]%
        {siddiqui2022isum}
\bibfield{author}{\bibinfo{person}{Tarique Siddiqui}, \bibinfo{person}{Saehan Jo}, \bibinfo{person}{Wentao Wu}, \bibinfo{person}{Chi Wang}, \bibinfo{person}{Vivek Narasayya}, {and} \bibinfo{person}{Surajit Chaudhuri}.} \bibinfo{year}{2022}\natexlab{}.
\newblock \showarticletitle{ISUM: Efficiently compressing large and complex workloads for scalable index tuning}. In \bibinfo{booktitle}{\emph{Proceedings of the 2022 International Conference on Management of Data}}. \bibinfo{pages}{660--673}.
\newblock


\bibitem[Siddiqui and Wu(2024)]%
        {siddiqui2024ml}
\bibfield{author}{\bibinfo{person}{Tarique Siddiqui} {and} \bibinfo{person}{Wentao Wu}.} \bibinfo{year}{2024}\natexlab{}.
\newblock \showarticletitle{ML-Powered Index Tuning: An Overview of Recent Progress and Open Challenges}.
\newblock \bibinfo{journal}{\emph{ACM SIGMOD Record}} \bibinfo{volume}{52}, \bibinfo{number}{4} (\bibinfo{year}{2024}), \bibinfo{pages}{19--30}.
\newblock


\bibitem[Stephens and Poess(2009)]%
        {stephens2009converting}
\bibfield{author}{\bibinfo{person}{John~M Stephens} {and} \bibinfo{person}{Meikel Poess}.} \bibinfo{year}{2009}\natexlab{}.
\newblock \showarticletitle{Converting TPC-H Query Templates to Use DSQGEN for Easy Extensibility}. In \bibinfo{booktitle}{\emph{Performance Evaluation and Benchmarking: First TPC Technology Conference, TPCTC 2009, Lyon, France, August 24-28, 2009, Revised Selected Papers 1}}. Springer, \bibinfo{pages}{99--115}.
\newblock


\bibitem[Stillger et~al\mbox{.}(2001)]%
        {Stillger_VLDB01}
\bibfield{author}{\bibinfo{person}{Michael Stillger}, \bibinfo{person}{Guy~M. Lohman}, \bibinfo{person}{Volker Markl}, {and} \bibinfo{person}{Mokhtar Kandil}.} \bibinfo{year}{2001}\natexlab{}.
\newblock \showarticletitle{{LEO}: {DB2}'s Learning Optimizer}. In \bibinfo{booktitle}{\emph{Proceedings of the 27th International Conference on Very Large Data Bases (VLDB)}}. \bibinfo{pages}{19--28}.
\newblock


\bibitem[Sun and Wang(2022)]%
        {sun2022toward}
\bibfield{author}{\bibinfo{person}{Hao Sun} {and} \bibinfo{person}{Taiyi Wang}.} \bibinfo{year}{2022}\natexlab{}.
\newblock \showarticletitle{Toward Causal-Aware RL: State-Wise Action-Refined Temporal Difference}.
\newblock \bibinfo{journal}{\emph{arXiv preprint arXiv:2201.00354}} (\bibinfo{year}{2022}).
\newblock


\bibitem[Sutton and Barto(2018)]%
        {sutton2018reinforcement}
\bibfield{author}{\bibinfo{person}{Richard~S Sutton} {and} \bibinfo{person}{Andrew~G Barto}.} \bibinfo{year}{2018}\natexlab{}.
\newblock \bibinfo{booktitle}{\emph{Reinforcement learning: An introduction}}.
\newblock \bibinfo{publisher}{MIT press}.
\newblock


\bibitem[Towers et~al\mbox{.}(2024)]%
        {towers2024gymnasium}
\bibfield{author}{\bibinfo{person}{Mark Towers}, \bibinfo{person}{Ariel Kwiatkowski}, \bibinfo{person}{Jordan Terry}, \bibinfo{person}{John~U Balis}, \bibinfo{person}{Gianluca De~Cola}, \bibinfo{person}{Tristan Deleu}, \bibinfo{person}{Manuel Goulao}, \bibinfo{person}{Andreas Kallinteris}, \bibinfo{person}{Markus Krimmel}, \bibinfo{person}{Arjun KG}, {et~al\mbox{.}}} \bibinfo{year}{2024}\natexlab{}.
\newblock \showarticletitle{Gymnasium: A standard interface for reinforcement learning environments}.
\newblock \bibinfo{journal}{\emph{arXiv preprint arXiv:2407.17032}} (\bibinfo{year}{2024}).
\newblock


\bibitem[Trummer et~al\mbox{.}(2019)]%
        {Trummer_SIGMOD19}
\bibfield{author}{\bibinfo{person}{Immanuel Trummer}, \bibinfo{person}{Junxiong Wang}, \bibinfo{person}{Deepak Maram}, \bibinfo{person}{Samuel Moseley}, \bibinfo{person}{Saehan Jo}, {and} \bibinfo{person}{Joseph Antonakakis}.} \bibinfo{year}{2019}\natexlab{}.
\newblock \showarticletitle{SkinnerDB: Regret-Bounded Query Evaluation via Reinforcement Learning}. In \bibinfo{booktitle}{\emph{Proceedings of the 2019 ACM SIGMOD International Conference on Management of Data}}. \bibinfo{pages}{1153--1170}.
\newblock


\bibitem[Valentin et~al\mbox{.}(2000)]%
        {valentin2000db2}
\bibfield{author}{\bibinfo{person}{Gary Valentin}, \bibinfo{person}{Michael Zuliani}, \bibinfo{person}{Daniel~C Zilio}, \bibinfo{person}{Guy Lohman}, {and} \bibinfo{person}{Alan Skelley}.} \bibinfo{year}{2000}\natexlab{}.
\newblock \showarticletitle{DB2 advisor: An optimizer smart enough to recommend its own indexes}. In \bibinfo{booktitle}{\emph{Proceedings of 16th International Conference on Data Engineering (Cat. No. 00CB37073)}}. IEEE, \bibinfo{pages}{101--110}.
\newblock


\bibitem[Van~Hasselt et~al\mbox{.}(2016)]%
        {van2016deep}
\bibfield{author}{\bibinfo{person}{Hado Van~Hasselt}, \bibinfo{person}{Arthur Guez}, {and} \bibinfo{person}{David Silver}.} \bibinfo{year}{2016}\natexlab{}.
\newblock \showarticletitle{Deep reinforcement learning with double q-learning}. In \bibinfo{booktitle}{\emph{Proceedings of the AAAI conference on artificial intelligence}}, Vol.~\bibinfo{volume}{30}.
\newblock


\bibitem[Wang et~al\mbox{.}(2024)]%
        {Wang_ICEBE24}
\bibfield{author}{\bibinfo{person}{Jiangqing Wang}, \bibinfo{person}{Yinglun Hu}, \bibinfo{person}{Chong Sun}, \bibinfo{person}{Huili Zhang}, {and} \bibinfo{person}{Chongwei Ruan}.} \bibinfo{year}{2024}\natexlab{}.
\newblock \showarticletitle{Joint Recommendation of Materialized Views and Indexes Based on Reinforcement Learning}. In \bibinfo{booktitle}{\emph{Proceedings of the 21st IEEE International Conference on e-Business Engineering (ICEBE)}}. \bibinfo{pages}{212--217}.
\newblock


\bibitem[Wang and Yoneki(2024)]%
        {wang2024ia2}
\bibfield{author}{\bibinfo{person}{Taiyi Wang} {and} \bibinfo{person}{Eiko Yoneki}.} \bibinfo{year}{2024}\natexlab{}.
\newblock \showarticletitle{IA2: Leveraging Instance-Aware Index Advisor with Reinforcement Learning for Diverse Workloads}. In \bibinfo{booktitle}{\emph{Proceedings of the 4th Workshop on Machine Learning and Systems}}. \bibinfo{pages}{10--17}.
\newblock


\bibitem[Wang et~al\mbox{.}(2022)]%
        {wang2022wetune}
\bibfield{author}{\bibinfo{person}{Zhaoguo Wang}, \bibinfo{person}{Zhou Zhou}, \bibinfo{person}{Yicun Yang}, \bibinfo{person}{Haoran Ding}, \bibinfo{person}{Gansen Hu}, \bibinfo{person}{Ding Ding}, \bibinfo{person}{Chuzhe Tang}, \bibinfo{person}{Haibo Chen}, {and} \bibinfo{person}{Jinyang Li}.} \bibinfo{year}{2022}\natexlab{}.
\newblock \showarticletitle{Wetune: Automatic discovery and verification of query rewrite rules}. In \bibinfo{booktitle}{\emph{Proceedings of the 2022 International Conference on Management of Data}}. \bibinfo{pages}{94--107}.
\newblock


\bibitem[Whang(1987)]%
        {whang1987index}
\bibfield{author}{\bibinfo{person}{Kyu-Young Whang}.} \bibinfo{year}{1987}\natexlab{}.
\newblock \showarticletitle{Index selection in relational databases}.
\newblock In \bibinfo{booktitle}{\emph{Foundations of Data Organization}}. \bibinfo{publisher}{Springer}, \bibinfo{pages}{487--500}.
\newblock


\bibitem[Yan et~al\mbox{.}(2021)]%
        {yan2021index}
\bibfield{author}{\bibinfo{person}{Yu Yan}, \bibinfo{person}{Shun Yao}, \bibinfo{person}{Hongzhi Wang}, {and} \bibinfo{person}{Meng Gao}.} \bibinfo{year}{2021}\natexlab{}.
\newblock \showarticletitle{Index selection for NoSQL database with deep reinforcement learning}.
\newblock \bibinfo{journal}{\emph{Information Sciences}}  \bibinfo{volume}{561} (\bibinfo{year}{2021}), \bibinfo{pages}{20--30}.
\newblock


\bibitem[Yang et~al\mbox{.}(2022)]%
        {Yang_SIGMOD22}
\bibfield{author}{\bibinfo{person}{Zongheng Yang}, \bibinfo{person}{Wei{-}Lin Chiang}, \bibinfo{person}{Sifei Luan}, \bibinfo{person}{Gautam Mittal}, \bibinfo{person}{Michael Luo}, {and} \bibinfo{person}{Ion Stoica}.} \bibinfo{year}{2022}\natexlab{}.
\newblock \showarticletitle{Balsa: Learning a Query Optimizer Without Expert Demonstrations}. In \bibinfo{booktitle}{\emph{Proceedings of the 2022 ACM SIGMOD International Conference on Management of Data}}. \bibinfo{pages}{931--944}.
\newblock


\bibitem[Yu et~al\mbox{.}(2022)]%
        {Yu_PVLDB22}
\bibfield{author}{\bibinfo{person}{Xiang Yu}, \bibinfo{person}{Chengliang Chai}, \bibinfo{person}{Guoliang Li}, {and} \bibinfo{person}{Jiabin Liu}.} \bibinfo{year}{2022}\natexlab{}.
\newblock \showarticletitle{Cost-based or Learning-based? A Hybrid Query Optimizer for Query Plan Selection}.
\newblock \bibinfo{journal}{\emph{Proc. VLDB Endowment}} \bibinfo{volume}{15}, \bibinfo{number}{13} (\bibinfo{year}{2022}), \bibinfo{pages}{3924--3936}.
\newblock


\bibitem[Yu et~al\mbox{.}(2020)]%
        {Yu_ICDE20}
\bibfield{author}{\bibinfo{person}{Xiang Yu}, \bibinfo{person}{Guoliang Li}, \bibinfo{person}{Chengliang Chai}, {and} \bibinfo{person}{Nan Tang}.} \bibinfo{year}{2020}\natexlab{}.
\newblock \showarticletitle{Reinforcement Learning with Tree-LSTM for Join Order Selection}. In \bibinfo{booktitle}{\emph{Proceedings of the 36th IEEE International Conference on Data Engineering (ICDE)}}. \bibinfo{pages}{1297--1308}.
\newblock


\bibitem[Zhang et~al\mbox{.}(2019)]%
        {Zhang_SIGMOD19}
\bibfield{author}{\bibinfo{person}{Ji Zhang}, \bibinfo{person}{Yu Liu}, \bibinfo{person}{Ke Zhou}, {and} \bibinfo{person}{Guoliang Li}.} \bibinfo{year}{2019}\natexlab{}.
\newblock \showarticletitle{An End-to-End Automatic Cloud Database Tuning System Using Deep Reinforcement Learning}. In \bibinfo{booktitle}{\emph{Proceedings of the 2019 ACM SIGMOD International Conference on Management of Data}}. \bibinfo{pages}{415--432}.
\newblock


\bibitem[Zhou et~al\mbox{.}(2024)]%
        {zhou2024breaking}
\bibfield{author}{\bibinfo{person}{Wei Zhou}, \bibinfo{person}{Chen Lin}, \bibinfo{person}{Xuanhe Zhou}, {and} \bibinfo{person}{Guoliang Li}.} \bibinfo{year}{2024}\natexlab{}.
\newblock \showarticletitle{Breaking It Down: An In-Depth Study of Index Advisors}.
\newblock \bibinfo{journal}{\emph{Proceedings of the VLDB Endowment}} \bibinfo{volume}{17}, \bibinfo{number}{10} (\bibinfo{year}{2024}), \bibinfo{pages}{2405--2418}.
\newblock


\bibitem[Zhou et~al\mbox{.}(2021)]%
        {zhou2021learned}
\bibfield{author}{\bibinfo{person}{Xuanhe Zhou}, \bibinfo{person}{Guoliang Li}, \bibinfo{person}{Chengliang Chai}, {and} \bibinfo{person}{Jianhua Feng}.} \bibinfo{year}{2021}\natexlab{}.
\newblock \showarticletitle{A learned query rewrite system using monte carlo tree search}.
\newblock \bibinfo{journal}{\emph{Proceedings of the VLDB Endowment}} \bibinfo{volume}{15}, \bibinfo{number}{1} (\bibinfo{year}{2021}), \bibinfo{pages}{46--58}.
\newblock


\bibitem[Zhu et~al\mbox{.}(2023)]%
        {Zhu_PVLDB23}
\bibfield{author}{\bibinfo{person}{Rong Zhu}, \bibinfo{person}{Wei Chen}, \bibinfo{person}{Bolin Ding}, \bibinfo{person}{Xingguang Chen}, \bibinfo{person}{Andreas Pfadler}, \bibinfo{person}{Ziniu Wu}, {and} \bibinfo{person}{Jingren Zhou}.} \bibinfo{year}{2023}\natexlab{}.
\newblock \showarticletitle{Lero: A Learning-to-Rank Query Optimizer}.
\newblock \bibinfo{journal}{\emph{Proc. VLDB Endowment}} \bibinfo{volume}{16}, \bibinfo{number}{6} (\bibinfo{year}{2023}), \bibinfo{pages}{1466--1479}.
\newblock


\end{thebibliography}


\end{document}